\documentclass[iop]{emulateapj}

\newcommand{\msun}{$\rm M_{\sun}$}
\newcommand{\rsun}{$\rm R_{\sun}$}
\newcommand{\rearth}{${\rm R_{\earth}} $}
\newcommand{\mearth}{$\rm M_{\earth}$}
\newcommand{\hho}{H$_2$O}
\newcommand{\hh}{H$_2$}
\newcommand{\chhhh}{CH$_4$}

\newcommand{\degree}{$^\circ$}

\newcommand{\zeroth}{$0^{\rm th}$}
\newcommand{\first}{$1^{\rm st}$}
\newcommand{\second}{$2^{\rm nd}$}

\newcommand{\fourth}{$4^{\rm th}$}
\newcommand{\e}{${\rm e^{-}}$}
\newcommand{\persecond}{${\rm s^{-1}}$}
\newcommand{\perpixel}{${\rm pixel^{-1}}$}
\newcommand{\divideoot}{{\tt divide-oot}}
\newcommand{\modelramp}{{\tt model-ramp}}

\newcommand{\chisq}{$\chi^2$}
\newcommand{\calwf}{{\tt calwf3}}
\newcommand{\rp}{$R_{p}$}
\newcommand{\rs}{$R_{\star}$}
\newcommand{\flt}{{\tt flt}}

\newcommand{\edit}[1]{{#1}}

\begin{document}

\shorttitle{WFC3 Observations of GJ1214b}
\shortauthors{Berta et al.}

\title{The Flat Transmission Spectrum of the Super-Earth GJ1214b from Wide Field Camera 3 on the Hubble Space Telescope}

\author{Zachory~K.~Berta\altaffilmark{1}, David~Charbonneau\altaffilmark{1},  Jean-Michel~D\'esert\altaffilmark{1}, Eliza~Miller-Ricci~Kempton\altaffilmark{2,8}, Peter~R.~McCullough\altaffilmark{3,4}, Christopher~J.~Burke\altaffilmark{5},  Jonathan~J.~Fortney\altaffilmark{2},  Jonathan~Irwin\altaffilmark{1}, Philip~Nutzman\altaffilmark{2},  Derek~Homeier\altaffilmark{6,7}}
\email{zberta@cfa.harvard.edu}
\altaffiltext{1}{Harvard-Smithsonian Center for Astrophysics, 60 Garden St.,
  Cambridge, MA 02138, USA}
\altaffiltext{2}{Department of Astronomy and Astrophysics, University of California, Santa Cruz, CA 95064, USA}
\altaffiltext{3}{Space Telescope Science Institute, 3700 San Martin Drive, Baltimore, MD 21218, USA}
\altaffiltext{4}{Smithsonian Astrophysical Observatory, 60 Garden St., Cambridge, MA 02138, USA}
\altaffiltext{5}{SETI Institute/NASA Ames Research Center, M/S 244-30, Moffett Field, CA 94035, USA}
\altaffiltext{6}{Centre de Recherche Astrophysique de Lyon, UMR 5574, CNRS, Universit\'e de Lyon, \'Ecole Normale Sup\'erieure de Lyon, 46 All\'ee d'Italie, F-69364 Lyon Cedex 07, France}
\altaffiltext{7}{Institut f\"ur Astrophysik, George-August-Universit\"at, Friedrich-Hund-Platz 1, 37077 G\"ottingen, Germany}
\altaffiltext{8}{Sagan Fellow}

\begin{abstract}

Capitalizing on the observational advantage offered by its tiny M dwarf host, we present HST/WFC3 grism measurements of the transmission spectrum of the super-Earth exoplanet GJ1214b. These are the first published WFC3 observations of a transiting exoplanet atmosphere. After correcting for a ramp-like instrumental systematic, we achieve nearly photon-limited precision in these observations, finding the transmission spectrum of GJ1214b to be flat between 1.1 and 1.7 \micron. Inconsistent with a cloud-free solar composition atmosphere at $8.2\sigma$, the measured achromatic transit depth most likely implies a large mean molecular weight for GJ1214b's outer envelope. A dense atmosphere rules out bulk compositions for GJ1214b that explain its large radius by the presence of a very low density gas layer surrounding the planet. High-altitude clouds can alternatively explain the flat transmission spectrum, but they would need to be optically thick up to 10 mbar or consist of \edit{particles with a range of sizes approaching 1 \micron~in diameter}.

\end{abstract}

\section{Introduction} \label{sec:introduction}

With a radius of 2.7 \rearth~and a mass of 6.5 \mearth, the transiting planet GJ1214b \citep{charbonneau.2009.stnls} is a member of the growing population of exoplanets whose masses and radii are known to be between those of Earth and Neptune \citep[see][]{leger.2009.tefcsmvcfswmr,batalha.2011.kfrpk,lissauer.2011.cpsllptk, winn.2011.stns}. Among these exoplanets, most of which exhibit such shallow transits that they require ultra-precise space-based photometry simply to  detect the existence of their transits, GJ1214b is unique. The diminutive 0.21 \rsun~radius of its M dwarf stellar host means GJ1214b exhibits a large 1.4\% transit depth, and the system's proximity (13 pc) means the star is bright enough \edit{in the near infrared} ($H=9.1$) that follow-up observations to study the planet's atmosphere are currently feasible. In this work, we exploit this observational advantage and present new measurements of the planet's atmosphere, which bear upon models for its interior composition and structure.

According to theoretical studies \citep{seager.2007.mrse,rogers.2010.tpol1,nettelmann.2011.tesmts1}, GJ1214b's 1.9 g cm$^{-3}$ bulk density is high enough to require a larger ice or rock core fraction than the solar system ice giants but far too low to be explained with an entirely Earth-like composition. \citet{rogers.2010.tpol1} have proposed three general scenarios consistent with GJ1214b's large radius, where the planet could (i) have accreted and maintained a nebular \hh/He envelope atop an ice and rock core, (ii) consist of a rocky planet with an \hh-rich envelope that formed by recent outgassing, or (iii) contain a large \edit{fraction of water} in its interior  surrounded by a dense \hh-depleted, \hho-rich atmosphere. Detailed thermal evolution calculations by \citet{nettelmann.2011.tesmts1} disfavor this last model on the basis that it would require unreasonably large bulk water-to-rock ratios, arguing for at least a partial \hh/He~envelope, albeit one that might be heavily enriched in \hho~relative to the primordial nebula.

By measuring GJ1214b's transmission spectrum, we can empirically constrain the mean molecular weight of the planet's atmosphere, thus distinguishing among these possibilities. When the planet passes in front of its host M dwarf,  a small fraction of the star's light passes through the upper layers of the planet's atmosphere before reaching us; the planet's transmission spectrum is then manifested in variations of the transit depth as a function of wavelength. The amplitude of the transit depth variations $\Delta D(\lambda)$ in the transmission spectrum scale as $n_{H}\times2HR_{p}/R_{\star}^2$, where \edit{$n_H$ is set by the opacities involved and can be 1-10 for strong absorption features}, $H$ is the atmospheric scale height, \rp~is the planetary radius, and \rs~is the stellar radius \citep[e.g.][]{seager.2000.ttsdegpt,brown.2001.tsdegpa, hubbard.2001.tegpt}. Because the scale height $H$ is inversely proportional to the mean molecular weight $\mu$ of the atmosphere, the amplitude of features seen in the planet's transmission spectrum places strong constraints on the possible values of $\mu$ and, in particular, the hydrogen/helium content of the atmosphere \citep{miller-ricci.2009.assdbhha}.

Indeed, detailed radiative transfer simulations of GJ1214b's atmosphere \citep{miller-ricci.2010.nats1} show that a solar composition, \hh-dominated atmosphere ($\mu=2.4$) would show depth variations of roughly $0.1\%$ between 0.6 and 10 \micron, while the features in an \hho-dominated atmosphere ($\mu = 18$) would be an order of magnitude smaller. While the latter of these is likely too small to detect directly with current instruments, the former is at a level that has regularly been measured with the Hubble Space Telescope (HST) in the transmission spectra of hot Jupiters \citep[e.g.][]{charbonneau.2002.depa,pont.2008.dahep0mts1wh,sing.2011.hsttse1hahonws}.

Spectroscopic observations by \citet{bean.2010.gtsse1} with the Very Large Telescope found the transmission spectrum of GJ1214b to be featureless between 0.78-1.0 \micron, down to an amplitude that would rule out cloud-free \hh-rich atmospheric models. Broadband Spizer Space Telescope photometric transit measurements at 3.6 and 4.5 \micron~by \citet{desert.2011.oemasg} showed a flat spectrum consistent with \citet{bean.2010.gtsse1}, as did high-resolution spectroscopy with NIRSPEC between 2.0 and 2.4 \micron~by \citet{crossfield.2011.hdnts1}. Intriguingly, the transit depth in $K$-band (2.2 \micron) was measured from CFHT by \citet{croll.2011.btss1smmwa} to be 0.1\% deeper than at other wavelengths, which would imply a \hh-rich atmosphere, in apparent contradiction to the other studies. 

These seemingly incongruous observations could potentially be brought into agreement if GJ1214b's atmosphere were \hh-rich but significantly depleted in \chhhh~\citep{crossfield.2011.hdnts1, miller-ricci-kempton.2011.ac1pc}. In such a scenario, the molecular features that remain (predominantly \hho) would fit the CFHT measurement, but be unseen by the NIRSPEC and Spitzer observations. Explaining the flat VLT spectrum in this context would then require a \edit{broadband} haze to smooth the spectrum at shorter wavelengths \citep[see][]{miller-ricci-kempton.2011.ac1pc}. New observations by \citet{bean.2011.ontsspgfema} covering 0.6-0.85 \micron~and 2.0-2.3 \micron~were again consistent with a flat spectrum, but they still could not directly speak to this possibility of a methane-depleted, \hh-rich atmosphere with optically scattering hazes. 

Here, we present a new transmission spectrum of GJ1214b spanning 1.1 to 1.7 \micron, using the infrared slitless spectroscopy mode on the newly installed Wide Field Camera 3 (WFC3) aboard the Hubble Space Telescope (HST). Our WFC3 observations directly probe the predicted strong 1.15 and 1.4 \micron~water absorption features in GJ1214b's atmosphere \citep[][]{miller-ricci.2010.nats1} and provide a stringent constraint on the \hh~content of GJ1214b's atmosphere that is robust to non-equilibrium methane abundances and hence a definite test of the \chhhh-depleted hypothesis. The features probed by WFC3 are the same features that define the $J$ and $H$ band windows in the telluric spectrum, and cannot be observed from the ground. 

Because this is the first published analysis of WFC3 observations of a transiting exoplanet, we include a detailed discussion of the performance of WFC3 in this observational regime and the systematic effects that are inherent to the instrument. Recent work on WFC3's predecessor NICMOS \citep{burke.2010.notjx, gibson.2011.lnts1gxcemf} has highlighted the importance of characterizing instrumental systematics when interpreting exoplanet results from HST observations. 

This paper is organized as follows: we describe our observations in \S\ref{sec:observations}, our method for extracting spectrophotometric light curves from them in \S\ref{sec:datareduction}, and our analysis of these light curves in \S\ref{sec:analysis}.  We present the resulting transmission spectrum and discuss its implications for GJ1214b's composition in \S\ref{sec:discussion}, and conclude in \S\ref{sec:conclusions}.

\section{Observations} \label{sec:observations}

We observed three transits of GJ1214b on UT 2010 October 8, 2011 March 28, and 2011 July 23 with the G141 grism on WFC3's infrared channel (HST Proposal \#GO-12251, P.I. = Z. Berta), obtaining simultaneous multiwavelength spectrophotometry of each transit between 1.1 and 1.7 \micron. WFC3's IR channel consists of a $1024\times1024$ pixel Teledyne HgCdTe detector with a 1.7 \micron~ cutoff that can be paired with any of 15 filters or 2 low-resolution grisms \citep{dressel.2010.wfcihv}. Each exposure is compiled from multiple non-destructive readouts and can consist of either the full array or a concentric, smaller subarray.

Each visit consisted of four 96 minute long HST orbits, each containing 45 minute gaps due to Earth occultations. Instrumental overheads between the occultations are dominated by serial downloads of the WFC3 image buffer, during which all science images are transferred to the telescope's solid state recorder. This buffer can hold only two 16-readout, full-frame IR exposures before requiring a download, which takes 6 minutes. Exposures cannot be started nor stopped during a buffer download, so parallel buffer downloads are impossible for short exposures.

Subject to these constraints and the possible readout sequences, we maximized the number of photons detected per orbit while avoiding saturation by gathering exposures using the $512\times512$ subarray with the RAPID NSAMP=7 readout sequence, for an effective integration time of 5.971 seconds  per exposure. With this setting, four 12-exposure batches, separated by buffer downloads, were gathered per orbit resulting in an integration efficiency of 10\%. Although the brightest pixel in the \first~order spectrum reaches 78\% of saturation during this exposure time, the WFC3's multiple non-destructive readouts enable the flux within each pixel to be estimated before the onset of significant near-saturation nonlinearities.

\begin{figure}[t] 
   \centering
   \includegraphics[width=\columnwidth]{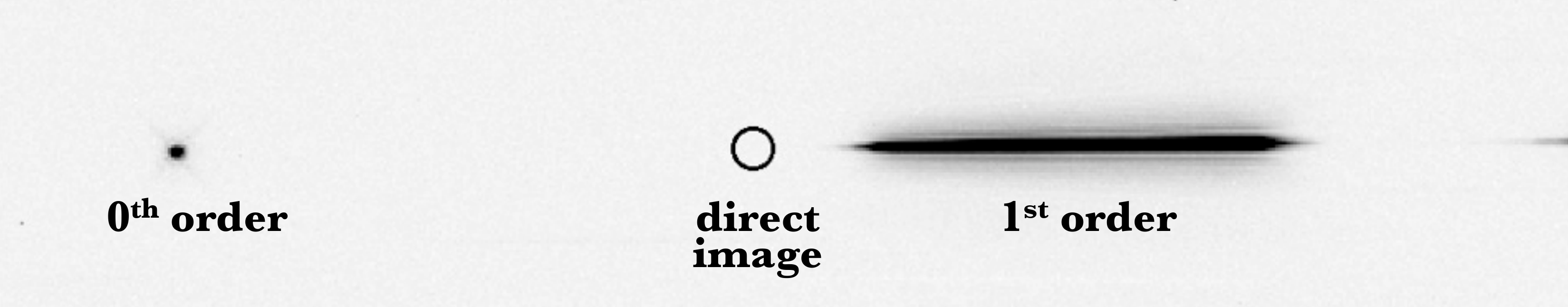} 
   \caption{A 512x100 pixel cutout of a typical WFC3 G141 grism exposure of the star GJ1214.  The \zeroth~and \first~order spectra are labeled, and the start of the \second~order spectrum is visible on the right. The location of the star in the direct images (not shown here) is marked with a circle. }
   \label{fig:typicalgrism}
\end{figure}

A sub-region of a typical G141 grism image of GJ1214 is shown in Fig.~\ref{fig:typicalgrism}. The $512\times512$ subarray allows both the \zeroth~ and \first~order spectra to be recorded, and the \first~order spectrum to fall entirely within a single amplifier quadrant of the detector. The \first~order spectrum spans 150 pixels with a dispersion in the $x$-direction of 4.65 nm/pixel and a spatial full-width half maximum in the $y$-direction of 1.7 pixels (0.2"). The \zeroth~order spectrum is slightly dispersed by the grism's prism but is nearly a point source. Other stars are present in the subarray's $68'\times61'$ field of view, but are too faint to provide useful diagnostics of systematic trends that may exist in the data. For wavelength calibration, we gathered direct images in the F130N narrow-band filter; the direct images' position relative to the grism images is also shown in Fig.~\ref{fig:typicalgrism}. 

To avoid systematics from the detector flat-fields that have a quoted precision no better than 0.5\% \citep{pirzkal.2011.fgiwf}, the telescope was not dithered during any of the observations. We note that a technique called ``spatial scanning'' has been proposed to decrease the overheads for bright targets with WFC3, where the telescope nods {\em during} an exposure to smear the light along the cross-dispersion direction, thus increasing the time to saturation \citep{mccullough.2011.si}. We did not use this mode of observation as it was not yet tested at the time our program was initiated.

\begin{figure}[t]
   \centering
   \includegraphics[width=3.5in]{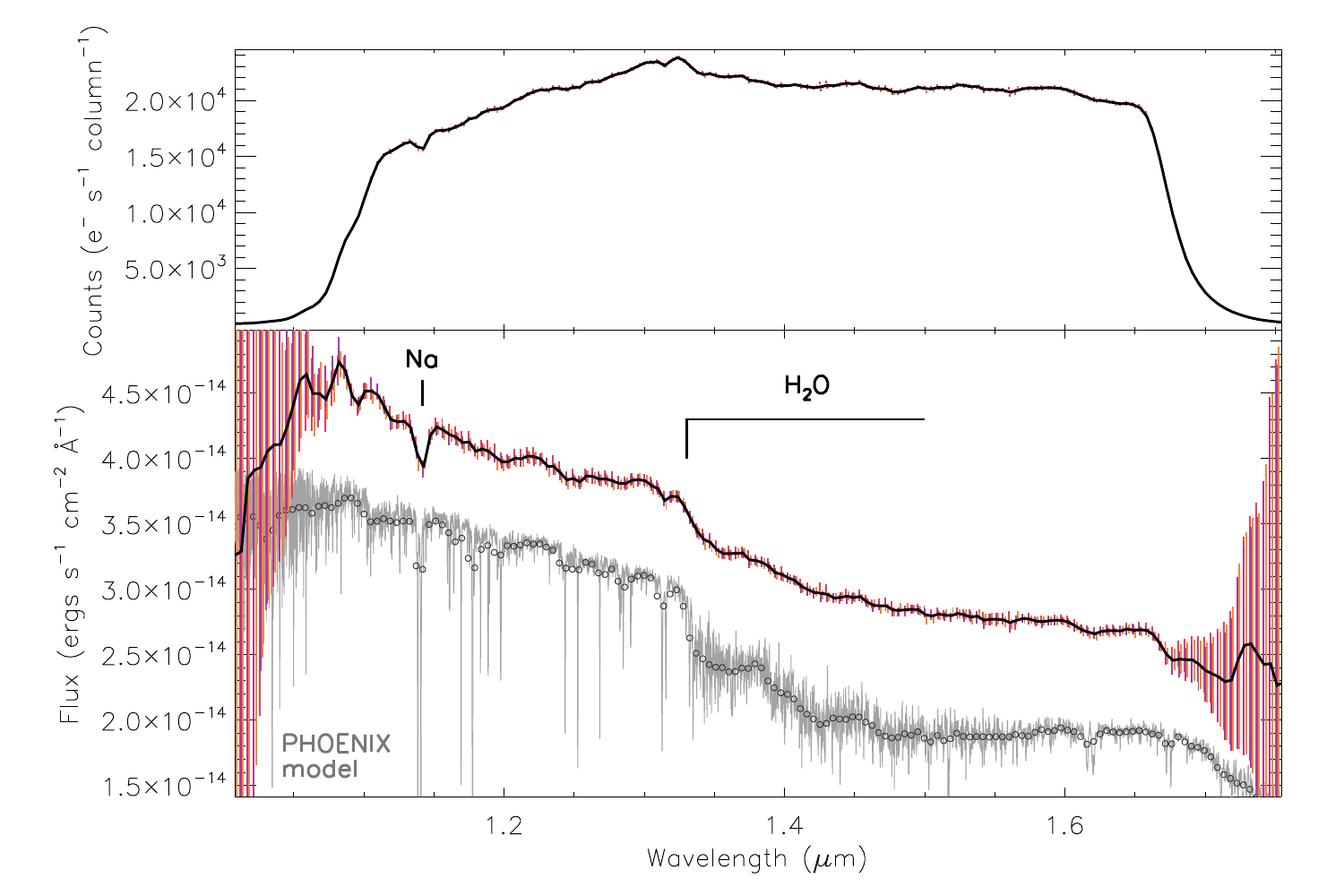} 
   \caption{The mean out-of-transit extracted spectrum of GJ1214 ({\em black line}) from all three HST visits, shown before ({\rm top}) and after ({\rm bottom}) flux calibration. Individual extracted spectra from each visit are shown with their $1\sigma$ uncertainties ({\em color error bars}). For comparison, the integrated flux from the PHOENIX model atmosphere used to calculate the stellar limb darkening (see \S\ref{sec:limbdarkening}) is shown ({\em gray lines}) offset for clarity and binned to the WFC3 pixel scale ({\em gray circles}).  }
    \label{fig:spectrum}
\end{figure}

\section{Data Reduction} \label{sec:datareduction}

The Python/PyRAF software package aXe was developed to extract spectra from slitless grism observations with WFC3 and other Hubble instruments \citep{kummel.2009.ssdes}, but it is optimized for extracting large numbers of spectra from full frame dithered grism images. To produce relative spectrophotometric measurements of our single bright source, we opted to create our own extraction pipeline that prioritizes precision in the time domain. We outline the extraction procedure below.

Through the extraction, we use calibrated 2-dimensional images, the ``\flt'' outputs from WFC3's \calwf~pipeline. For each exposure, \calwf~performs the following steps: flag detector pixels with the appropriate data quality (DQ) warnings, estimate and remove bias drifts using the reference pixels, subtract dark current, determine count rates and identify cosmic rays by fitting a slope to the non-destructive reads, correct for photometric non-linearity (properly accounting for the signal accumulation before the initial ``zeroth'' read), and apply gain calibration. The resulting images are measured in \e s$^{-1}$ and contain per pixel uncertainty estimates based on a detector model \citep{kim-quijano.2009.wmh}. We note that \calwf~does {\em not} apply flat-field corrections when calibrating grism images; proper wavelength-dependent flat-fielding for slitless spectroscopy requires wavelength-calibrating individual sources and \calwf~does not perform this task. 

\subsection{Interpolating over Cosmic Rays}\label{sec:cosmics}
\calwf~identifies cosmic rays that appear partway through an exposure by looking for deviations from a linear accumulation of charge among the non-destructive readouts, but it can not identify cosmic rays that appear between the zeroth and first readout. We supplement \calwf's cosmic ray identifications by also flagging any pixel in an individual exposure that is $>6\sigma$ above the median of that pixel's value in all other exposures as a cosmic ray. Through all three visits (576 exposures), a total of 88 cosmic rays were identified within the extraction box for the \first~order spectra.

For each exposure, we spatially interpolate over cosmic rays. Near the \first~order spectrum, the pixel-to-pixel gradient of the point spread function (PSF) is typically much shallower along the dispersion direction than perpendicular to it, so we use only horizontally adjacent pixels when interpolating to avoid errors in modeling the sharp cross-dispersion falloff.

\subsection{Identifying Continuously Bad Pixels}\label{sec:badpixels}

We also mask any pixels that are identified as ``bad detector pixels'' (DQ=4), ``unstable response'' (DQ=32), ``bad or uncertain flat value'' (DQ=512). We found that only these DQ flags affected the photometry in a pixel by more than $1\sigma$. Other flags may have influenced the pixel photometry, but did so below the level of the photon noise. In the second visit, we also identified one column of the detector ($x$ = 625 in physical pixels\footnote{For ease of comparison with future WFC3 analyses, throughout this paper we quote all pixel positions in physical units as interpreted by SAOImage DS9, where the bottom left pixel of a full-frame array would be $(x,y)$ = (1,1).}) whose light curve exhibited a dramatically different systematic variation than did light curves from any of the other columns. This column was coincident with an unusually low-sensitivity feature in the flat-field, and we hypothesize that the flat-field is more uncertain in this column than in neighboring columns. We masked all pixels in that column as bad.

We opt not to interpolate over these continuously bad pixels. Because they remained flagged throughout the duration of each visit, we simply give these pixels zero weight when extracting 1D spectra from the images. This allows us to keep track of the actual number of photons recorded in each exposure so we can better assess our predicted photometric uncertainties.

\begin{figure}[t] 
   \centering
   \includegraphics[width=\columnwidth]{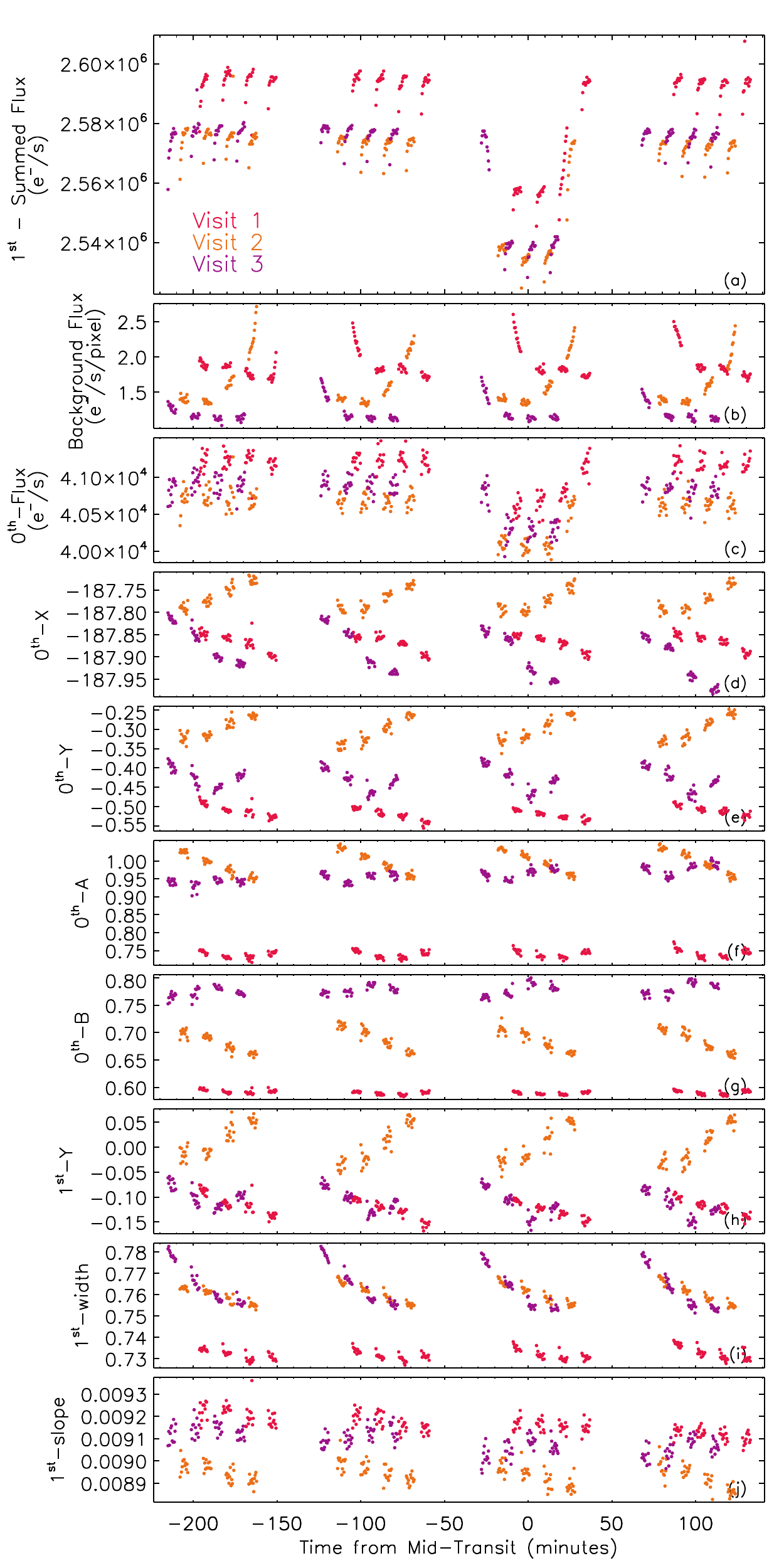} 
   \caption{Extracted properties of the \zeroth~and \first~order spectra as a function of time, including (a) the summed \first~order photometric light curves; (b) the estimated sky background level; (c-g) the total flux, $x$ and $y$ position (measured relative to the reference pixel), and Gaussian widths in the $x$ and $y$ directions of the \zeroth~order image;  and (h-j) the $y$ offset, cross-dispersion width, and slope of the \first~order spectrum. All three visits are shown and are denoted by the color of the  symbols. The 45 minute gaps in each time series are due to Earth occultations, the 6 minute gaps are due to the WFC3 buffer downloads.}
   \label{fig:timeseries}

\end{figure}

\subsection{Background Estimation}
In addition to the target, WFC3 also detects light from the diffuse sky background, which comes predominantly from zodiacal light and Earth-shine, and must be subtracted. We draw conservative masks around all sources that are visible in each visit's median image, including GJ1214 and its electronics cross-talk artifact \citep[see][]{viana.2010.wtc}. We exclude these pixels, as well as all pixels that have any DQ warning flagged. Then, to estimate the sky background in each exposure, we scale a master WFC3 grism sky image \citep{kummel.2011.miwggg} to match the remaining 70-80\% of the pixels in each exposure and subtract it. We find typical background levels of $1-3$ \e~\persecond~\perpixel, that vary smoothly within orbits and throughout visits as shown in \edit{Fig.~\ref{fig:timeseries} (panel b)}. As a test, we also estimated the background level from a simple mean of the unmasked pixels; the results were unchanged.

\subsection{Inter-pixel Capacitance}
The normal calibration pipeline does not correct for the inter-pixel capacitance (IPC) effect, which effectively couples the flux recorded in adjacent pixels at about the 1\% level \citep{mccullough.2008.icpd}. We correct this effect with a linear deconvolution algorithm \citep{mccullough.2008.icpd, hilbert.2011.iccmmo}, although we find it makes little difference to the final results.

\subsection{Extracting the Zeroth Order Image}\label{sec:extractingzeroth}
The \zeroth~order image can act as a diagnostic for tracking changes in the telescope pointing and in the shape of the instrumental PSF. We select a $10\times10$ pixel box around the \zeroth~order image and fit a 2D Gaussian to it with the $x$ position, $y$ position, size in the $x$ direction, size in the $y$ direction, and total flux allowed to vary (5 parameters). 

Time series of the \zeroth~order x and y positions, sizes in both directions, and total flux are shown for all three visits in \edit{Fig.~\ref{fig:timeseries} (panels c-g)}. Thanks to the dispersion by the grism's prism, the Gaussian is typically 20\% wider in the x direction than in the y direction. Even though the throughput of the \zeroth~order image is a factor of 60 lower than the \first~order spectrum, the transit of GJ1214b is readily apparent in the \zeroth-order flux time series.

\subsection{Extracting the First Order Spectrum}\label{sec:extractfirst}
To extract the first order spectra, we first determine the position of GJ1214 in the direct image, which serves as a reference position for defining the trace and wavelength calibration of the \first~order spectrum. We adopt the mean position GJ1214 in all of the direct images as the reference position, which we measure using the same method as in extracting the \zeroth~order image in \S\ref{sec:extractingzeroth}. The measured $(x,y)$ reference positions for the first, second, and third visits are (498.0, 527.5), (498.6, 531.1), and (498.9, 527.1) in physical pixels.

Once the reference pixel for a visit is known, we use the coefficients stored in the WFC3/G141 aXe configuration file\footnote{The aXe configuration file {\tt WFC3.IR.G141.V2.0.conf} is available through \url{http://www.stsci.edu/hst/wfc3/}} \citep[][]{kuntschner.2009.wsp1cgg},  to determine the geometry of the \first~order trace and cut out a 30 pixel tall extraction box centered on the trace. Within this extraction box, we use the wavelength calibration coefficients to determine the average wavelength of light that will be illuminating each pixel. We treat all pixels in the same column as having the same effective wavelength; given the spectrum's 0.5\degree~tilt from to the $x$ axis, errors introduced by this simplification are negligible. 

\citet{kuntschner.2008.gcwg} used flat-fields taken through all narrow-band filters available on WFC3/IR to construct a flat-field ``cube'' where each pixel contains 4 polynomial coefficients that describe its sensitivity as a function of wavelength\footnote{{{\tt WFC3.IR.G141.flat.2.fits}, through the same URL}}. We use this flat-field cube to construct a color-dependent flat based on our estimate of the effective wavelength illuminating each pixel, and divide each exposure by it. WFC3 wavelength calibration and flat-fielding is described in detail in the aXe manual \citep{kummel.2010.umv}.

To calculate 1D spectra from the flat-fielded images, we sum all the unmasked pixels within the extraction box over the $y$-axis. To estimate the uncertainty in each spectral channel, we first construct a per-pixel uncertainty model that includes photon noise from the source and sky as well as 22 \e~of read noise, and sum these uncertainties, in quadrature, over the $y$-axis. We do not use the \calwf-estimated uncertainties; they include a term propagated from the uncertainty in the nonlinearity correction that, while appropriate for absolute photometry, would not be appropriate for relative photometry. In each exposure, there are typically $1.2\times10^5$ \e~per single-pixel spectral channel and a total of $1.5\times10^7$ \e~in the entire spectrum. Fig.~\ref{fig:timeseries}  \edit{(panel a)} shows the extracted spectra summed over all wavelengths as a function of time, the ``white'' light curve.

For diagnostics' sake, we also measure the geometrical properties of the \first~order spectra in each exposure. We fit 1D Gaussians the cross-dispersion profile in each column of the spectrum and take the median Gaussian width among all the columns as a measurement of the PSF's width. We fit a line to the location of the Gaussian peaks in all the columns, taking the intercept and the slope of that line as an estimate of the $y$-offset and tilt of the spectrum on the detector. Time series of these parameters are shown in Fig.~\ref{fig:timeseries} \edit{(panels h-j)}.

\subsection{Flux Calibration}

For the sake of display purposes only (see Fig.~\ref{fig:spectrum}), we flux calibrate each visit's median, extracted, 1D spectrum. Here we have interpolated over all bad pixels within each visit (contrary to the discussion in the \S\ref{sec:badpixels}), and plotted the weighted mean over all three visits. The calibration uncertainty for the G141 sensitivity curve \citep{kuntschner.2011.rfcwggg} is quoted to be 1\%.

\subsection{Times of Observations}
For each exposure, we extract the {\tt EXPSTART} keyword from the science header, which is the Modified Julian Date at the start of the exposure. We correct this to the mid-exposure time using the {\tt EXPTIME} keyword, and convert it to the Barycentric Julian Date in the Barycentric Dynamical Time standard using the code provided by \citet{eastman.2010.abtmahbjd}.

\begin{deluxetable*}{rrrrrrrrrrrr}
\tablecaption{\label{tab:spectrum}White Light Curves from WFC3/G141}
\tabletypesize{\normalsize}
\tablecolumns{12}
\tablehead{\colhead{Time\tablenotemark{a}} & \colhead{Relative Flux\tablenotemark{b}} & \colhead{Uncertainty} & \colhead{Sky} & \colhead{\zeroth-X\tablenotemark{c}}& \colhead{\zeroth-Y\tablenotemark{c}} & \colhead{\zeroth-A\tablenotemark{d}}& \colhead{\zeroth-B\tablenotemark{d}}& \colhead{\first-Y\tablenotemark{c}}&  \colhead{\first-B\tablenotemark{d}}  & \colhead{\first-Slope\tablenotemark{e}}  & \colhead{Visit} \\  \colhead{$({\rm BJD_{TDB})}$} & & & \colhead{(\e/s)} & \colhead{(pix)}& \colhead{(pix)}& \colhead{(pix)}& \colhead{(pix)} & \colhead{(pix)}& \colhead{(pix)}& \colhead{(pix/pix)}  } 
\startdata
2455478.439980 & 0.99381 & 0.00031 &   1.9546 & -187.830 &   -0.474 &    0.790 &    0.613 &   -0.075 &   0.7484 &  0.00921 & 1 \\ 
2455478.440270 & 0.99713 & 0.00031 &   1.9938 & -187.839 &   -0.481 &    0.792 &    0.616 &   -0.082 &   0.7490 &  0.00925 & 1 \\ 
2455478.440559 & 0.99787 & 0.00031 &   1.9718 & -187.846 &   -0.484 &    0.789 &    0.615 &   -0.076 &   0.7486 &  0.00914 & 1 \\ 
2455478.440848 & 0.99958 & 0.00032 &   1.8808 & -187.827 &   -0.490 &    0.792 &    0.614 &   -0.087 &   0.7476 &  0.00917 & 1 \\ 
2455478.441138 & 0.99989 & 0.00032 &   1.9379 & -187.844 &   -0.490 &    0.785 &    0.612 &   -0.085 &   0.7492 &  0.00921 & 1 \\ 
 &  &  &    & ... &   &     &     &  &   &  &  \\ 
\enddata

\tablenotetext{a}{Mid-exposure time.}
\tablenotetext{b}{Normalized to the median flux level of the out-of-transit observations in each visit.}
\tablenotetext{c}{Position measured relative to the Gaussian center of each visit's direct image.}
\tablenotetext{d}{Gaussian width of the \zeroth~or \first~order spectra in the horizontal (A) or vertical (B) direction.}
\tablenotetext{e}{Slope of the \first~order spectrum.}

\tablecomments{This table is published in its entirety in the electronic edition of the Astrophysical Journal. A portion is shown here for guidance concerning its form and content.}
\end{deluxetable*}

\section{Analysis} \label{sec:analysis}
In this section we describe our method for estimating parameter uncertainties (\S\ref{sec:estimatingdistributions}) and our strategy for modeling GJ1214's stellar limb darkening (\S\ref{sec:limbdarkening}). Then, after identifying the dominant systematics in WFC3 light curves (\S\ref{sec:systematics}) and describing a method to correct them (\S\ref{sec:correctingsystematics}), we present our fits to the light curves, both summed over wavelength (\S\ref{sec:lcanalysis}) and spectroscopically resolved (\S\ref{sec:scanalysis}). We also present a fruitless search for transiting satellite companions to GJ1214b (\S\ref{sec:moons}). 

\subsection{Estimating Parameter Distributions}\label{sec:estimatingdistributions}

Throughout our analysis, we fit different WFC3 light curves with models that have different sets of parameters, and draw conclusions from the inferred probability distributions of those parameters; this section describes our method for characterizing the posterior probability distribution for a set of parameters within a given model.

We use a Markov Chain Monte Carlo (MCMC) method with the Metropolis-Hastings algorithm to explore the posterior probability density function (PDF) of the model parameters. This Bayesian technique allows us to sample from (and thus infer the shape of) the probability distribution of a model's parameters given both our data and our prior knowledge about the parameters \citep[for reviews, see][]{ford.2005.quoep,gregory.2005.bldapscawms, hogg.2010.darfmd}. Briefly, the algorithm starts a chain with an initial set of parameters ($\mathbf{M}_{j=0}$) and generates a trial set of parameters ($\mathbf{M'}_{j+1}$) by perturbing the previous set. The ratio of posterior probability between the two parameter sets, given the data $\mathbf{D}$, is then calculated as 
\begin{equation}
\frac{P(\mathbf{M'}_{j+1} | \mathbf{D})}{P(\mathbf{M}_{j }| \mathbf{D})} = \frac{P(\mathbf{D} | \mathbf{M'}_{j+1})}{P(\mathbf{D} | \mathbf{M}_{j})}\times \frac{P( \mathbf{M'}_{j+1})}{P(\mathbf{M}_{j})}
\label{eq:bayes}
\end{equation}
where the first term (the ``likelihood'') accounts for the information that our data provide about the parameters and the second term (the ``prior'') specifies our externally conceived knowledge about the parameters. If a random number drawn from a uniform distribution between 0 and 1 is less than this probability ratio, then $\mathbf{M}_{j+1}$ is set to $\mathbf{M'}_{j+1}$; if not, then $\mathbf{M}_{j+1}$ reverts to $\mathbf{M}_{j}$. The process is iterated until $j$ is large, and the resulting chain of parameter sets is a fair sample from the posterior PDF and can be used to estimate confidence intervals for each parameter. 

To calculate the likelihood term in Eq.~\ref{eq:bayes}, we assume that each of the $N$ flux values $d_{i}$ is drawn from a uncorrelated Gaussian distribution centered on the model value $m_i$ with a standard deviation of $s\sigma_i$, where $\sigma_i$ is the theoretical uncertainty for the flux measurement based on the detector model and photon statistics and $s$ is a photometric uncertainty rescaling parameter. Calculation of the ratio in  Eq.~\ref{eq:bayes} is best done in logarithmic space for numerical stability, so we write the likelihood as
\begin{equation}
\ln P(\mathbf{D} | \mathbf{M}) = - N \ln s - \frac{1}{2s^2}\chi^2 + {\rm constant}
\end{equation}
where 
\begin{equation}
\chi^2 = \sum_{i=1}^{N}\left(\frac{d_i - m_i}{\sigma_i}\right)^2
\end{equation}
and we have only explicitly displayed terms that depend on the model parameters. Including $s$ as a model parameter is akin to rescaling the uncertainties by externally modifying $\sigma_i$ to achieve a reduced $\chi^2$ of unity, but enables the MCMC to fit for and marginalize over this rescaling automatically. Unless otherwise stated for specific parameters, we use non-informative (uniform) priors for the second term in Eq.~\ref{eq:bayes}. We use a Jeffreys prior on $s$ (uniform in $\ln s$) which is the least informative, although the results are practically indistinguishable from prior uniform in $s$. 

When generating each new trial parameter set $\mathbf{M'}_{j+1}$, we follow \citet{dunkley.2005.frmcmctcpe} and perturb every parameter at once, drawing the parameter jumps from a multivariate Gaussian with a covariance matrix that approximates that of the parameter distribution. Doing so allows the MCMC to move easily along the dominant linear correlations in parameter PDF, and greatly increases the efficiency of the algorithm. While this procedure may seem circular (if we knew the covariance matrix of the parameter distribution, why would we need to perform the MCMC?), the covariance matrix we use to generate trial parameters could be a very rough approximation to the true shape of the parameter PDF but still dramatically decrease the computation time necessary for the MCMC.

To obtain an initial guess for parameters ($\mathbf{M_{j=0}}$), we use the {\tt MPFIT} implementation \citep{markwardt.2009.nlfwm} of the Levenberg-Marquardt (LM) method to maximize $\ln P(\mathbf{M} | \mathbf{D})$. This would be identical to minimizing $\chi^2$ in the case of flat priors, but it can also include constraints from more informative priors. The LM fit also provides an estimate of the covariance matrix of the parameters, which is a linearization of the probability space near the best-fit. We use this covariance matrix estimate for generating trial parameters in the MCMC, \edit{and with it, achieve parameter acceptance rates of 10-40\% throughout the following sections.} As expected, when fitting models with flat priors and linear or nearly-linear parameters (where the PDF should well-described by a multivariate Gaussian), the LM covariance matrix is identical to that ultimately obtained from the MCMC \citep[see][for further discussion]{sivia.1996.dabt}.

MCMC chains are run until they contain $1.25\times10^5$ points. The first 1/5 of the points are ignored as ``burn-in'', leaving $1\times10^5$ for parameter estimation. \edit{Correlation lengths for the parameters in the MCMC chains are indicated throughout the text; they are typically of order 10 points. A chain with such a correlation length effectively contains $1\times10^5/10 = 1\times10^4$  independent realizations of the posterior PDF.} We quote confidence intervals that exclude the upper and lower 16\% of the marginalized distribution for each parameter (i.e. the parameter's central 68\% confidence interval), \edit{using all 1$\times$10$^5$ points in each chain}. 

\subsection{Modeling Stellar Limb Darkening}\label{sec:limbdarkening}

Accurate modeling of the WFC3 integrated and spectroscopic transit light curves requires careful consideration of the stellar limb-darkening (LD) behavior. GJ1214b's M4.5V stellar host is so cool that it exhibits weak absorption features due to molecular \hho. Because inferences of the planet's apparent radii from transit light curves depend strongly on the star's limb-darkening, which is clearly influenced by \hho~as an opacity source, inaccurate treatment of limb-darkening could potentially introduce spurious \hho~features into the transmission spectrum. 

If they were sufficiently precise, transit light curves alone could simultaneously constrain both the star's multiwavelength limb-darkening behavior and the planet's multiwavelength radii \citep[e.g.][]{knutson.2007.uslrp2}. For less precise light curves, it is common practice to fix the limb-darkening to a theoretically calculated law, even if this may underestimate the uncertainty in the planetary parameters \citep[see][]{burke.2007.xtjmcpmb,southworth.2008.hstepila}. Given the quality of our data, we adopt an intermediate solution where we allow the limb-darkening parameters to vary in our fits, but with a Gaussian prior centered on the theoretical values \citep[e.g.][]{bean.2010.gtsse1}.

We model the star GJ1214's limb-darkening behavior with a spherically symmetric {\tt PHOENIX} atmosphere \citep{hauschildt.1999.nmag3}, assuming stellar parameters of $T_{\rm eff} = 3026K$, $\log g = 5$, and [M/H] = 0 \citep{charbonneau.2009.stnls}. As shown in Fig.~\ref{fig:spectrum}, the integrated flux from the PHOENIX model is in good qualitative agreement with the low-resolution, calibrated WFC3 stellar spectrum of GJ1214. From this model, we calculate photon-weighted average intensity profiles for the integrated spectrum and for each of the individual wavelength bins, using the WFC3 grism sensitivity curve and the {\tt PHOENIX} model to estimate the photon counts. In the spherical geometry of the PHOENIX atmospheres the characterization of the actual limb (defined as $\mu=0$, see below) is not straightforward, as the model extends beyond the photosphere into the optically thin outer atmosphere. The result is an approximately exponentially declining intensity profile from the
outermost layers, that \citet{claret.2003.lssnmamss} found not to be easily reproduced by standard limb darkening laws for plane-parallel atmospheres. These authors suggest the use of ``quasi-spherical" models by ignoring the outer region. In an extension of this concept, we set the outer surface of the star to be where the intensity drops to $e^{-1}$ of the central intensity, and measure $\mu = \cos\theta$ (where $\theta$ is the emission angle relative to the line of sight) relative to that outer radius. 

We derive coefficients for a square-root limb-darkening law for each of these average intensity profiles using least-squares fitting. In this law, the intensity relative to the center of the star is given by
\begin{equation}
\frac{I(\mu)}{I(1)} = 1 - c(1-\mu) - d (1-\sqrt{\mu}),
\end{equation}
where $c$ and $d$ are the two coefficients of the fit. We chose a square-root law over the popular quadratic law because it gave noticeably better approximations to the {\tt PHOENIX} intensity profiles, while still having few enough free parameters that they can be partially inferred from the data. Indeed, \citet{van-hamme.1993.lcmbslc} found the square-root law to be generally preferable to other 2-parameter limb-darkening laws for late-type stars in the near-IR. The square-root law matches the theoretical intensity profile nearly as well as the full nonlinear 4-parameter law introduced by \cite{claret.2000.nlsamcl2t5ssg} for the models we use here.

\begin{figure}[t]
   \centering
   \includegraphics[width=\columnwidth]{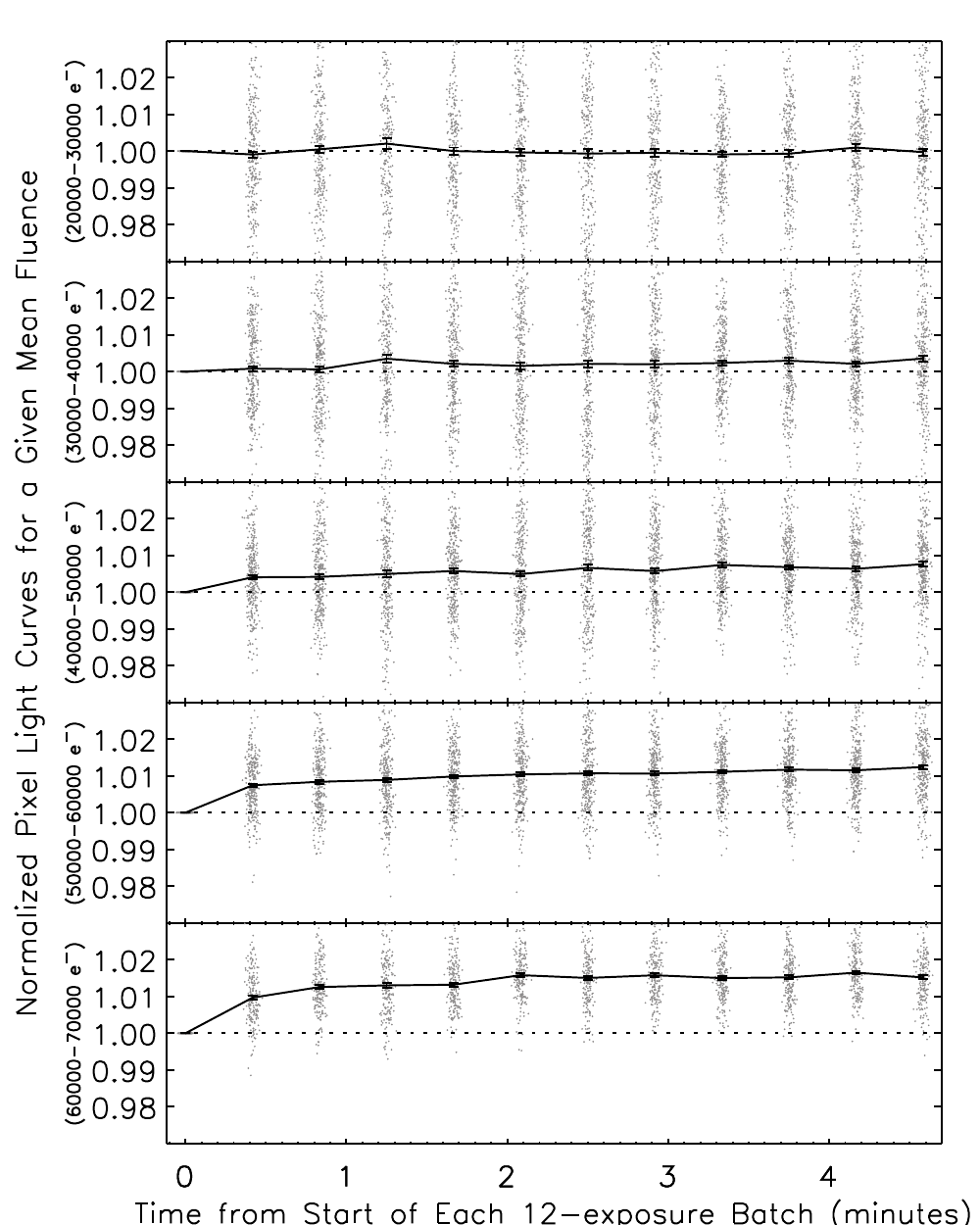} 
   \caption{Single-pixel light curves within each 12-exposure batch following a buffer download ({\em gray points}), shown for different mean pixel illuminations. Pixel light curves have been normalized to the first exposure within each batch, and plotted with small random horizontal offsets for clarity. Only data from the first HST visit, which exhibited the smallest pointing drifts (see Fig.~\ref{fig:timeseries}), are shown. Error bars show the mean and its standard error for each time point and each illumination. An exponential ramp begins is present in pixels with a mean recorded fluence greater than 30,000 and 40,000 \e~(50\% of the detector full well). Note, the nominal fluences quoted here do not include charge accumulated during detector flushing and initial readout (see text).}
    \label{fig:ramp}
   \end{figure}

\subsection{Light Curve Systematics}\label{sec:systematics}

The summed light curve shown in Fig.~\ref{fig:timeseries} \edit{(panel a)} exhibits non-astrophysical systematic trends. The most obvious of these are the sharply rising but quickly saturating ``ramp''-like features within each batch of 12 exposures between buffer downloads. To the eye, the ramps are very repeatable; the flux at the end of all batches asymptotes to nearly the same  level. The amplitude of the ramp is 0.4\% from start to finish for most batches, except for the first batch of each orbit, where the ramp is somewhat less pronounced. 

These ramps are reminiscent of those seen in high-cadence Spitzer light curves at 8 and 16 \micron~\citep[e.g.][]{deming.2006.siefep1, knutson.2007.dcep1, charbonneau.2008.biese1} which \citet{agol.2010.c1fftems} recently proposed  may be due to ``charge trapping'' within the detector pixels. In their toy model, charge traps within each pixel become filled throughout an exposure and later release the trapped charge on a finite timescale, thereby increasing the pixel's dark current in subsequent exposures. The model leads to exponential ramps when observing bright sources as the excess dark current increases sharply at first but slows its increase as the population of charge traps begins to approach steady state. We note this model also leads to after-images following strong exposures, i.e., persistence.

WFC3 has been known since its initial ground-testing to exhibit strong persistence behavior \citep{mccullough.2008.wtp, long.2010.wpmcutle}. \citet{smith.2008.tiphp} have proposed that persistence in 1.7 \micron~cutoff HgCdTe detectors like WFC3 is likely related to charge trapping. Measurements \citep{mccullough.2008.wtp} indicate that WFC3's persistence may be of the right order of magnitude (on $<1$ minute timescales) to supply the roughly 50 \e~\persecond~\perpixel~in the brightest pixels that would be necessary to explain the observed several millimagnitude ramp, although persistence levels and decay timescales can depend in complicated ways on the strength of previous exposures \citep[see][]{smith.2008.ciphp}. 

We were aware of this persistence issue before our observations and made an effort to control its effect on our light curves. When we planned the timing of the exposures, we attempted to make the illumination history of each pixel as consistent as possible from batch to batch and orbit to orbit. In practice, this means we gathered more direct images than necessary for wavelength calibration to delay some of the grism exposures. 

Whether or not the ramps are caused by the charge trapping mechanism, they are definitely dependent on the illumination that a pixel receives. To demonstrate this, we construct light curves for each individual pixel over the duration of every out-of-transit 12-exposure batch that follows a buffer download and normalize each of these pixel light curves to the first exposure in the batch. Fig.~\ref{fig:ramp} shows the normalized pixel light curves, grouped by their mean recorded fluence. Because it takes a finite time to read the subarray (0.8 seconds) and reset the full array (2.9 seconds), we note that each exposure actually collects 60\% more electrons than indicated by these nominal, recorded fluences \citep[see][]{long.2011.dtp}. The appearance of the ramp clearly becomes more pronounced for pixels that are more strongly exposed. 

Buried beneath the ramp features, the summed light curve exhibits subtler trends that appear mostly as orbit-long or visit-long slopes with a peak-to-peak variation of about 0.05\%.  These are perhaps caused by slow drifts in pointing and focus (telescope ``breathing'') interacting with sensitivity variations across the detector that are not perfectly corrected by the flat field. 

\subsection{Correcting for Systematics}\label{sec:correctingsystematics}

Fortunately, these systematics are extremely repeatable between orbits within a visit; we harness this fact when correcting for them. We divide the in-transit orbit of any photometric timeseries, either the white light curve or one of the spectroscopically resolved light curves, by a systematics correction template constructed from the two good out-of-transit orbits. This template is simply the weighted average of the fluxes in the out-of-transit orbits, evaluated at each exposure within an orbit. It encodes both variations in the effective sensitivity of the detector within an orbit and the mean out-of-transit flux level. 

When performing the division, we propagate the template uncertainty into the photometric uncertainty for each exposure, which typically increases it by a factor of $\sqrt{1 + 1/2} = 1.22$. This factor, although it may seem like an undesired degradation of the photometric precision, would inevitably propagate into measurements of the transit depth whether we performed this correction or not, since \rp/\rs~is always measured relative to the out-of-transit flux, which must at some point be inferred from the data.

Throughout this work, we refer to this process of dividing by the out-of-transit orbits as the \divideoot~method. Because each point in the single in-transit orbit is equally spaced in time between the two out-of-transit exposures being used to correct it, the \divideoot~method also naturally removes the 0.05\% visit-long slope seen in the raw photometry. As we show in \S\ref{sec:lcanalysis}, when applied to the white light curves, the \divideoot~treatment produces uncorrelated Gaussian residuals that have a scatter consistent with the predicted photon uncertainties. 

Unlike decorrelation techniques that have often been used to correct systematics in HST light curves, the \divideoot~method does not require knowing the relationship between measured photometry and the physical state of the camera. It does, however, strictly require the systematics to repeat over multiple orbits. The \divideoot~method would not work if the changes in the position, shape, and rotational angle of the \first-order spectrum were not repeated in the other orbits in a visit or if the cadence of the illumination were not nearly identical across orbits. \edit{In such cases, the Gaussian process method proposed by \citet{gibson.2011.gpfmisats} may be a useful alternative, and one that would appropriately account for the uncertainty involved in the systematics correction.}

\begin{figure*}[t] 
   \centering
   \includegraphics[width=\columnwidth]{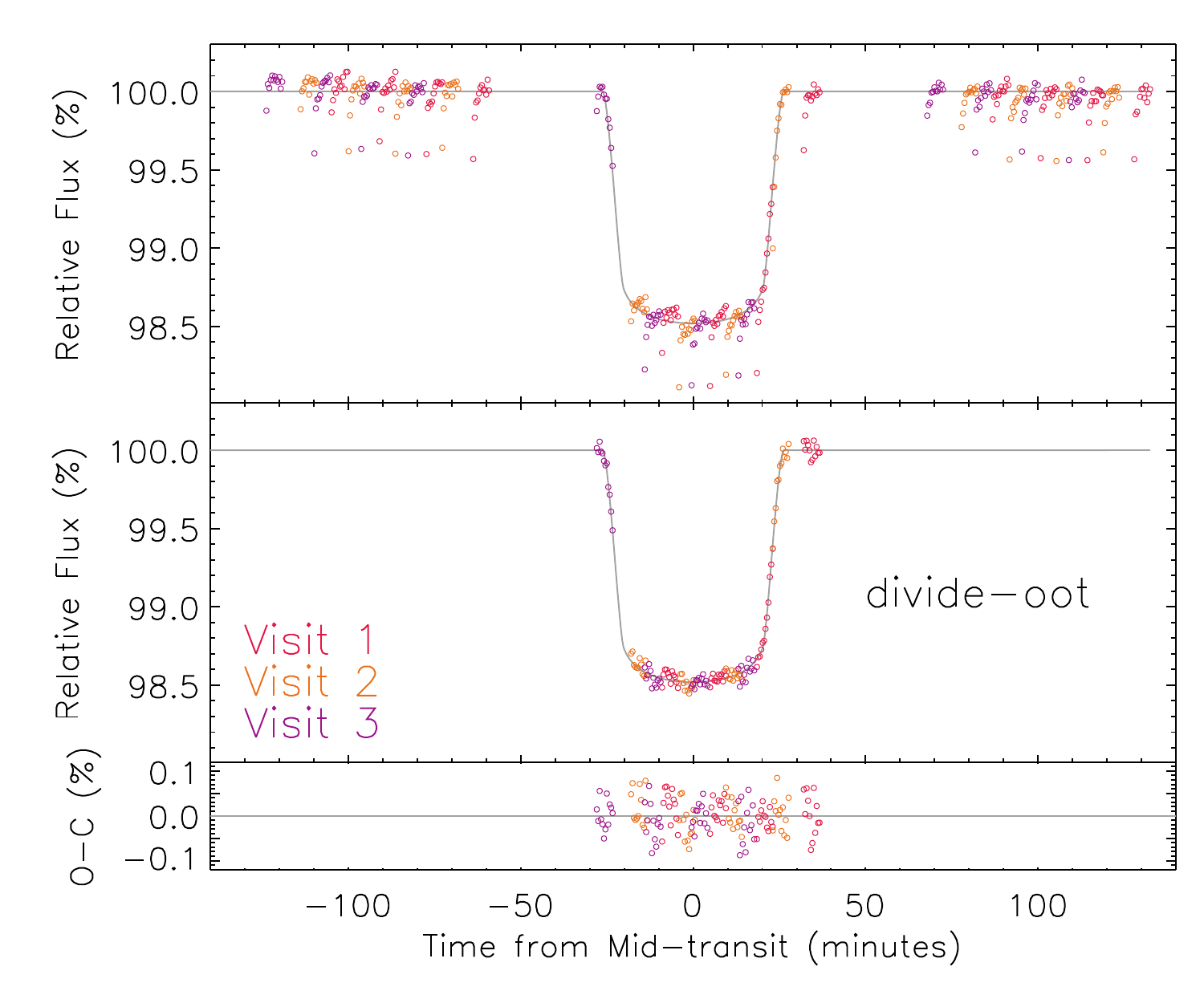} 
   \includegraphics[width=\columnwidth]{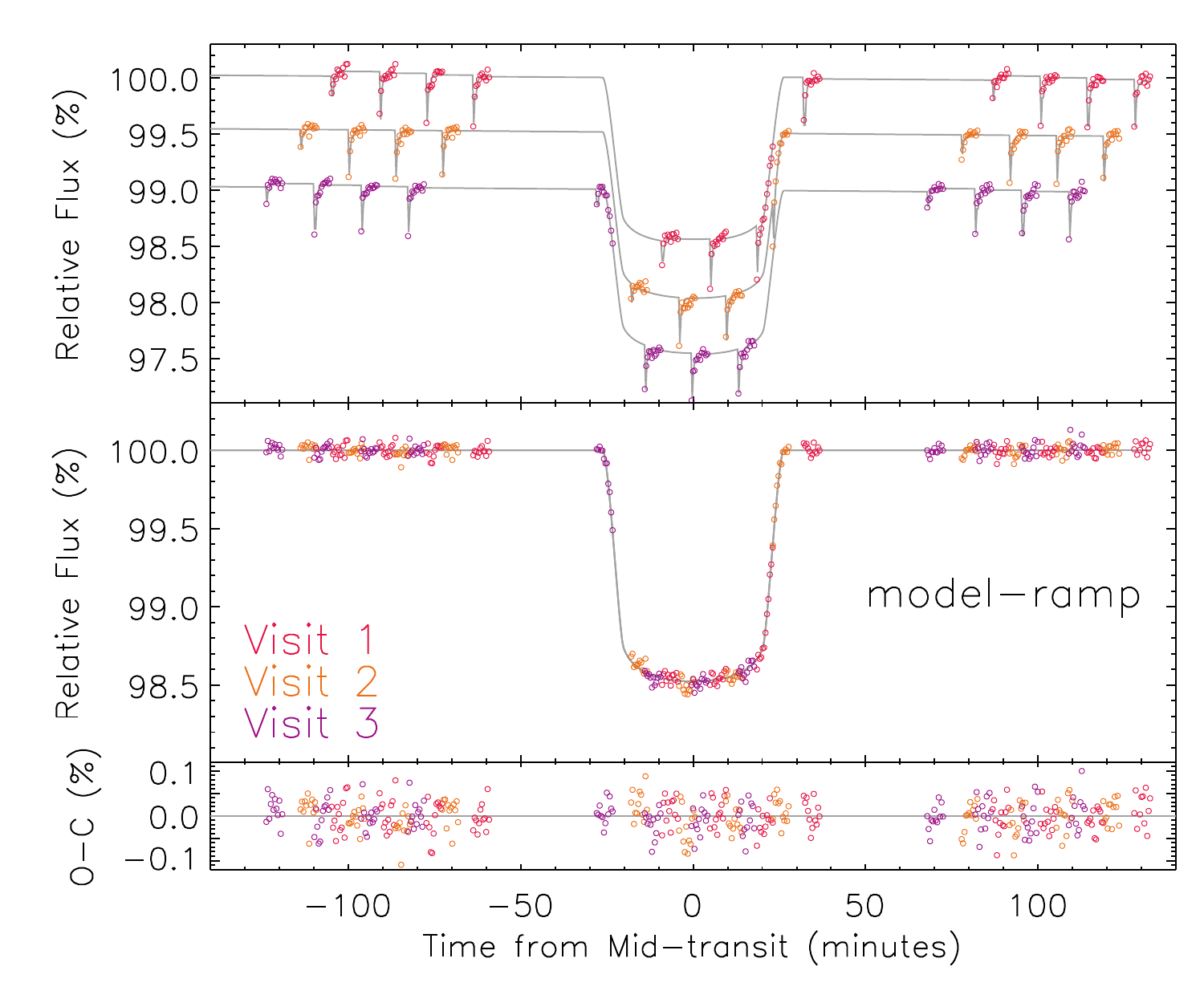}
   \caption{The white light curve of GJ1214b's transits before ({\it top panels}) and after ({\it middle panels}) removing the instrumental systematics using the \divideoot~({\em left}) and \modelramp~({\em right}, with offsets for clarity) methods described in \S\ref{sec:correctingsystematics} and \S\ref{sec:ramp}. A transit model that was fit to the \divideoot-corrected light curve, constrained to the values of $a/R_\star$ and $b$ used by \citet{bean.2010.gtsse1}, is shown ({\em gray lines}), along with residuals from this model ({\it bottom panels}). In the left panels, the out-of-transit orbits are not shown after the correction has been applied, because they contain no further information.}
   \label{fig:white}
   \end{figure*}

\subsection{White Light Curve Fits}\label{sec:lcanalysis}

Although the main scientific result of this paper is derived from the spectroscopic light curves presented in \S\ref{sec:scanalysis}, we also analyze the light curve summed over all wavelengths between 1.1 and 1.7 \micron. We use these white light curves to confirm the general system properties found in previous studies and quantitatively investigate the instrumental systematics.

We fit an analytic, limb-darkened transit light curve model \citep{mandel.2002.alcpts} to the \divideoot-corrected white light curves. Only the in-transit orbits were fit; after the \divideoot~correction, the two out-of-transit orbits contain no further information. Also, because the in-transit orbit's flux has already been normalized, we fix the out-of-transit flux level to unity in all the fits. Throughout, we fix the planet's period to  $P=1.58040481$ days and mid-transit time to $T_{c} = 2454966.525123$ ${\rm BJD_{TDB}}$ \citep{bean.2011.ontsspgfema}, the orbital eccentricity to $e=0$, and the stellar mass to 0.157\msun~\citep{charbonneau.2009.stnls}.

\subsubsection{Combined White Light Curve}\label{sec:combined}
First, we combine the three visits into a single light curve, as shown in Fig.~\ref{fig:white}, and fit for the following parameters: the planet-to-star radius ratio (\rp/\rs), the total transit duration between first and fourth contact ($t_{14}$), the stellar radius (\rs), and the two coefficients $c$ and $d$ of the square-root limb-darkening law\footnote{The square-root law is a special case of the 4-parameter law and straightforward to include in the \citet{mandel.2002.alcpts} model.}. Previous studies have found no significant transit timing variations for the GJ1214b system \citep{charbonneau.2009.stnls,sada.2010.rtse1,bean.2010.gtsse1,carter.2011.tlcpxsts1,desert.2011.oemasg,kundurthy.2011.ao1spesa,berta.2011.gsssvtsap,croll.2011.btss1smmwa}, so we fix the time of mid-transit for each visit to be that predicted by the linear ephemeris.

As in \citet{burke.2007.xtjmcpmb}, we use the parameters $t_{14}$ and \rs~to ensure quick convergence of the MCMC because correlations among these parameters are more linear than for the commonly fit impact parameter ($b$) and scaled semi-major axis ($a/R_{\star}$). Because nonlinear transformations between parameter pairs will deform the hypervolume of parameter space, we include a Jacobian term in the priors in Eq.~\ref{eq:bayes} to ensure uniform priors for the physical parameters $R_{p}$, $R_\star$, and $i$ \citep[see][for detailed discussions]{burke.2007.xtjmcpmb, carter.2008.aatlouc}. For the combined light curves, the influence of this term is practically negligible, but we include it for completeness. In the MCMC chains described in this section, all parameters have correlation lengths of 6-13 points.

Initially, we perform the fit with limb-darkening coefficients $c$ and $d$ without any priors from the {\tt PHOENIX} atmosphere model,  enforcing only that $0 < c+d < 1$, which ensures that the star is brighter at its center ($\mu=1$) than at its limb ($\mu = 0$). Interestingly, the quantity $(c/3 + d/5)$, which sets the integral of $I(\mu)$ over the stellar surface, defines the line along which $c$ and $d$ are most strongly correlated in the MCMC samples \citep[see also][]{irwin.2011.ljd4mebfmts}. For quadratic limb-darkening, the commonly quoted $2u_1+u_2$ combination \citep[][]{holman.2006.tlcpifctex} has the same physical meaning. The integral of $I(\mu)$ can be thought of as the increase in the central transit depth over that for a constant-intensity stellar disk, so it makes sense that it is well-constrained for nearly equatorial transiting systems like GJ1214b. Planets with higher impact parameters do not sample the full range of $0<\mu<1$ during transit, leading to correspondingly weaker limb-darkening constraints that can be derived from their light curves \citep[see][]{knutson.2011.stse4esvcdfv}. We quote confidence intervals for the linear combination $(c/3 + d/5)$ and one orthogonal to it in Table~\ref{tab:combinedfit}, along with rest of the parameters. 

Heartened by finding that when they are allowed to vary freely, our inferred white-light limb-darkening coefficients agree to $1\sigma$ to those derived using the {\tt PHOENIX} stellar model, we perform a second fit that includes the {\tt PHOENIX} models as informative priors. For this prior, we say $P(\mathbf{M})$ in Eq.~\ref{eq:bayes} is proportional to a Gaussian with $(c/3 + d/5) = 0.0892 \pm 0.018$ and $(c/5 - d/3) =-0.431 \pm  0.032$, which is centered on the {\tt PHOENIX} model. To set the $1\sigma$ widths of these priors, we start by varying the effective temperature of the star in the {\tt PHOENIX} model by its 130K uncertainty in either direction, and then double the width of the prior beyond this, to account for potential systematic uncertainties in the atmosphere model. The results from the fit with these LD priors are shown in Table~\ref{tab:combinedfit}.

\begin{figure}[t] 
   \centering
   \includegraphics[width=\columnwidth]{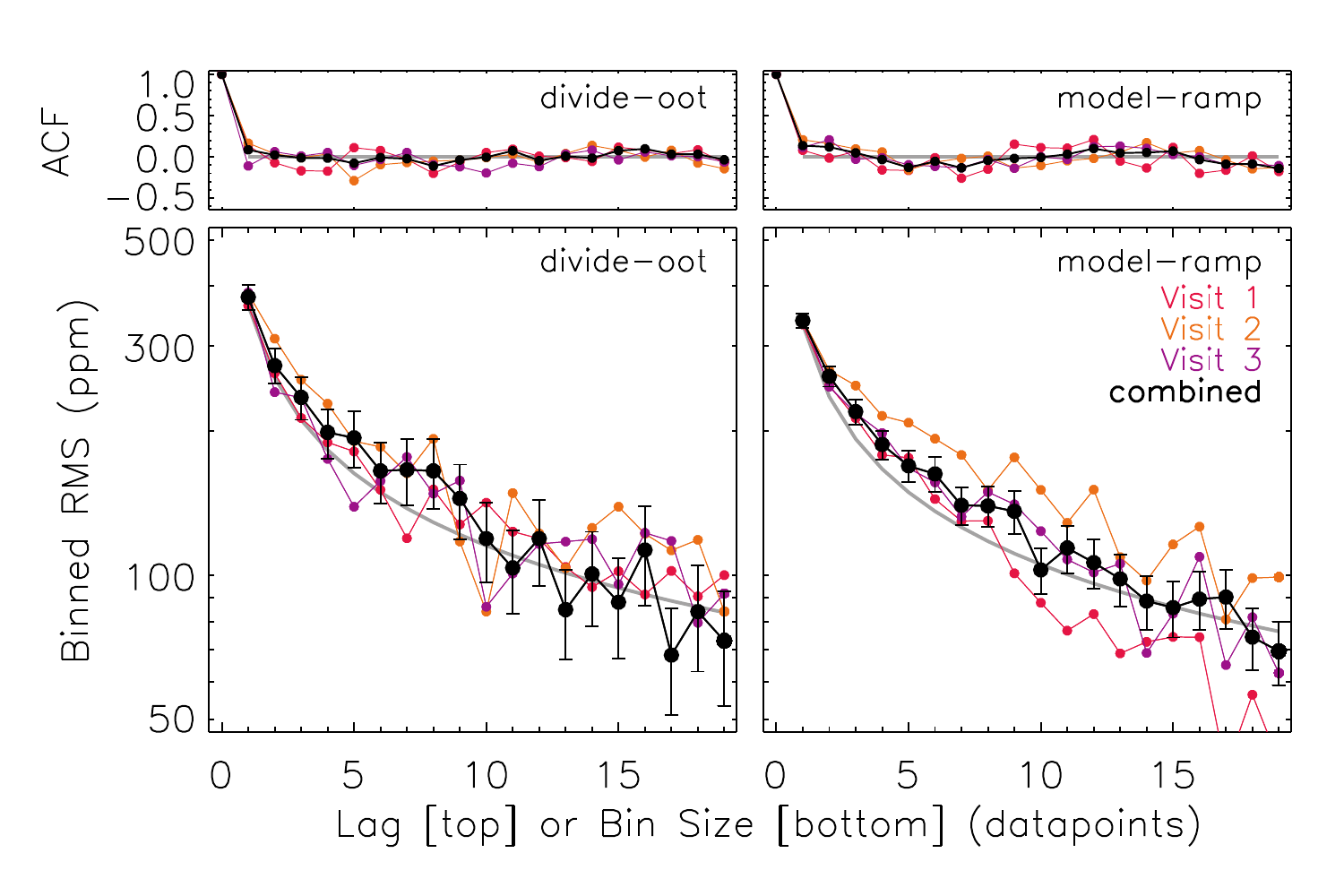} 
   \caption{For the transit model in Fig.~\ref{fig:white} and both types of systematics treatments, the autocorrelation function of the residuals (ACF; {\em top}) and the scatter in binned residuals as a function of bin size ({\em bottom}). The residuals from the combined light curve are shown ({\em black points}), as well as the individual visits ({\em colorful points}). The expectations from uncorrelated Gaussian noise (0 in the top, $\propto1/\sqrt{N}$ in the bottom) are overplotted ({\em dashed lines}).}
   \label{fig:residuals}
   \end{figure}

The photometric noise rescaling parameter $s$ is within 10\% of unity, implying that the \edit{376 ppm} achieved scatter in the combined white light curve can be quite well-explained from the known sources of uncertainty in the measurements, predominantly photon noise from the star. As shown in Fig.~\ref{fig:residuals}, for the \divideoot-corrected light curves, the autocorrelation function (ACF) of the residuals shows no evidence for time-correlated noise. Likewise, the scatter in binned \divideoot~residuals decreases as the square-root of the number of points in a bin, as expected for uncorrelated Gaussian noise. If there are uncorrected systematic effects remaining in the data, they are below the level of the photon noise over the time-scales of interest here. 

\begin{deluxetable}{lcc}
\tablecaption{\label{tab:combinedfit}Transit Parameters Inferred from the Combined White Light Curve}
\tabletypesize{\normalsize}
\tablecolumns{3}
\tablehead{\colhead{Parameter} & \colhead{No LD Prior} & \colhead{LD Prior\tablenotemark{a}}}
\startdata
$t_{14}$ (days)   &   $0.03624\pm0.00013$   &   $0.03620\pm0.00012$\\
$R_\star$ (\rsun)       &   $0.2014^{+0.0038}_{-0.0025}$   &   $0.201^{+0.004}_{-0.003}$\\
$a/R_\star$       &   $15.30^{+0.19}_{-0.29}$   &   $15.31^{+0.21}_{-0.29}$\\
$i$ (\degree)       &   $89.3\pm0.4$   &   $89.3^{+0.4}_{-0.3}$\\
$b $       &   $0.18^{+0.09}_{-0.11}$   &   $0.19^{+0.08}_{-0.11}$\\
$R_p/R_\star$\tablenotemark{b}       &   $0.1158^{+0.0007}_{-0.0006}$   &   $0.1160\pm0.0005$\\
$c/3 + d/5$       &   $0.096\pm0.008$   &   $0.095\pm0.007$\\
$c/5 - d/3$       &   $-0.52^{+0.22}_{-0.14}$   &   $-0.433\pm0.032$\\
\\
predicted RMS       &    337 ppm   &    337 ppm\\
achieved RMS       &    373 ppm   &    376 ppm\\
$s$       &   $1.12\pm0.07$   &   $1.12\pm0.07$\\
\enddata
\tablenotetext{a}{The Gaussian limb-darkening priors of $(c/3 + d/5) = 0.0892\pm 0.018$ and $c/5 - d/3 =-0.4306\pm 0.032$ were derived from {\tt PHOENIX} stellar atmospheres, as described in the text.}
\tablenotetext{b}{Confidence intervals on $R_p/R_\star$ do not include the $\Delta R_p/R_\star=0.0006$ systematic uncertainty due to stellar variability (see text).}
\end{deluxetable}

\subsubsection{Individual White Light Curves}\label{sec:individual}

To test for possible differences among our WFC3 visits, we fit each of the three \divideoot-corrected white light curves individually. In addition to \rp/\rs, $t_{14}$, and \rs, we also allow $\Delta T_{c}$ (the deviation of each visit's mid-transit time from the linear ephemeris) to vary freely. We allow $c$ and $d$ to vary, but enforce the same {\tt PHOENIX}-derived priors described in \S\ref{sec:combined}. 

Table~\ref{tab:visits} shows the results, which are consistent with each other and with other observations \citep[][]{charbonneau.2009.stnls,bean.2010.gtsse1,carter.2011.tlcpxsts1, kundurthy.2011.ao1spesa,berta.2011.gsssvtsap}. The three measured $\Delta T_{c}$'s show no evidence for transit timing variations.  \edit{The uncertainties for the parameters $t_{14}$, \rs, and $\Delta T_{c}$ are noticeably largest in the first visit; this is most likely because the first visit does not directly measure the timing of either \first~or \fourth~contact, on which these parameters strongly depend. Additionally, whereas the correlation lengths in the MCMC chains for these parameters in the two visits that do measure \first/\fourth~contact and for \rp/\rs, $c$, and $d$ in all three visits are small (10-30 points), the correlation lengths for $t_{14}$, \rs, and $\Delta T_{c}$ in the first visit are very large (300-400 points), indicating these weakly constrained parameters are poorly approximated by the MPFIT-derived covariance matrix. On account of the large correlation lengths for these parameters, we ran the MCMC for the first visit with a factor of 10 more points.} In each of the three visits, the uncertainty rescaling parameter $s$ is \edit{slightly above but} consistent with unity, \edit{indicating the photometric scatter is quite well explained by known sources of noise.}

\setlength{\tabcolsep}{0.02in} 
 \begin{deluxetable}{lccc}
\tablecaption{\label{tab:visits}Transit Parameters Inferred from Individual White Light Curves}
\tabletypesize{\normalsize}
\tablecolumns{4}
\tablehead{\colhead{Parameter} & \colhead{Visit 1} & \colhead{Visit 2} & \colhead{Visit 3}}
\startdata
$\Delta T_{c}$ (days)\tablenotemark{a}    &   $-0.0001^{+0.0012}_{-0.0020}$   &   $0.00028\pm0.00031$   &   $0.0002\pm0.0004$\\
$t_{14}$ (days)   &   $0.037^{+0.004}_{-0.003}$   &   $0.0357\pm0.0007$   &   $0.0369\pm0.0010$\\
$R_\star$ (\rsun)      &   $0.211^{+0.021}_{-0.014}$   &   $0.200^{+0.008}_{-0.006}$   &   $0.214^{+0.015}_{-0.011}$\\
$a/R_\star$       &   $14^{+1}_{-1}$   &   $15.4^{+0.5}_{-0.6}$   &   $14.4\pm0.9$\\
$i$ (\degree)       &   $88.9\pm0.7$   &   $89.2\pm0.6$   &   $88.5^{+0.8}_{-0.7}$\\
$b $       &   $0.27\pm0.17$   &   $0.21\pm0.15$   &   $0.38^{+0.13}_{-0.20}$\\
$R_p/R_\star$       &   $0.1164^{+0.0009}_{-0.0008}$   &   $0.1159^{+0.0011}_{-0.0009}$   &   $0.1175^{+0.0011}_{-0.0012}$\\
\\
predicted RMS       &    337 ppm   &    337 ppm   &    337 ppm\\
acheived RMS       &    343 ppm   &    360 ppm   &    366 ppm\\
$s$       &   $1.07^{+0.13}_{-0.11}$   &   $1.12^{+0.13}_{-0.11}$   &   $1.14^{+0.14}_{-0.11}$\\
\enddata
\tablenotetext{a}{\footnotesize Offset between the observed mid-transit time and that calculated from the linear ephemeris with $P=1.58040481$ and $T_c$=2454966.525123 ${\rm BJD_{TDB}}$.}
\end{deluxetable}

\subsubsection{Stellar Variability}
\begin{figure}[t] 
   \centering
   \includegraphics[width=\columnwidth]{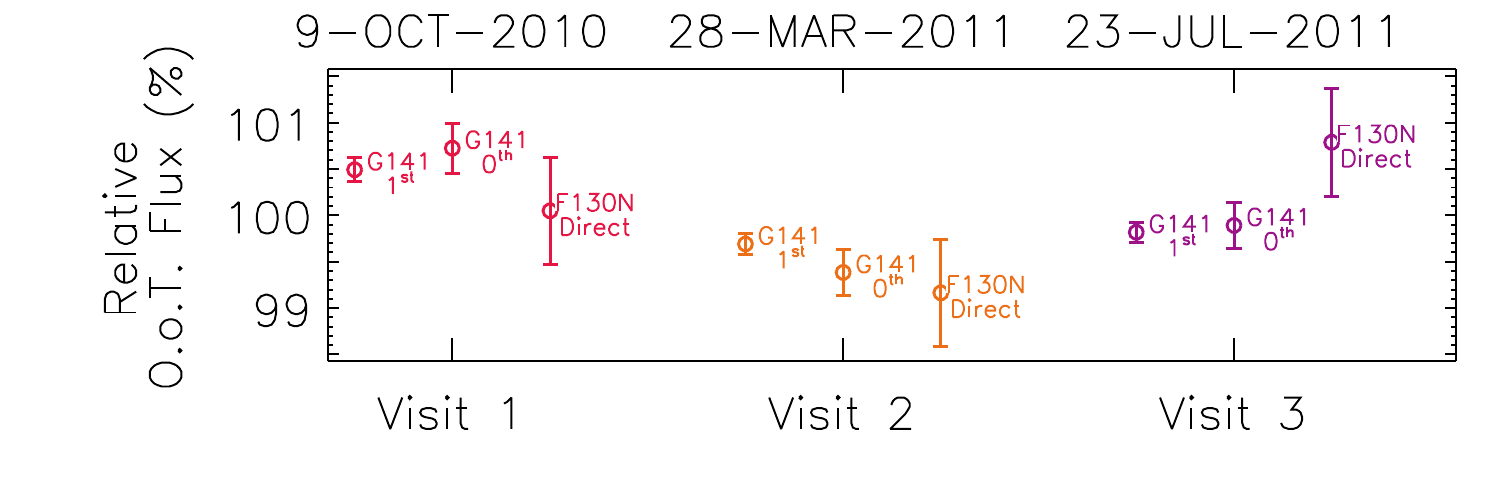} 
   \caption{The relative out-of-transit (O.o.T.) flux for each HST visit, measured independently from three different groups of images:  the summed \first-order spectrum, the \zeroth-order image, and the narrow-band direct image, each normalized to its mean. Error bars denote the standard deviation of the out-of-transit measurements within each visit; they do not include the 0.5\% uncertainty in the detector flat-field. \edit{The narrow-band measurements sample fewer photons, thus their larger uncertainties.}}
   \label{fig:variability}
\end{figure}

GJ1214 is known to be variable on 50-100 day timescales with an amplitude of 1\% in the MEarth bandpass \citep[715-1000 nm; see][]{charbonneau.2009.stnls, berta.2011.gsssvtsap}. To gauge the impact of stellar variability in the wavelengths studied here, we plot in Fig.~\ref{fig:variability} the relative out-of-transit flux as measured by our WFC3 data. For each HST visit, we have three independent measurements of this quantity: the F130N narrow-band direct image, the \zeroth-order spectrum, and the \first-order spectrum. Consistent variability over these measurements that sample different regions of the detector within each visit would be difficult to reproduce by instrumental effects, such as flat-fielding errors. In Fig.~\ref{fig:variability}, GJ1214 appears brighter in the first visit than in the last two visits, with an overall amplitude of variation of about 1\%.

This 1\% variability, if caused by unocculted spots on the stellar surface, should lead to variations in the inferred planet-to-star radius ratio on the order of $\Delta R_p/R_\star=0.0006$ \citep{berta.2011.gsssvtsap}. This is larger than the formal error on \rp/\rs~from the combined white light curve (Tab.~\ref{tab:combinedfit}), and must be considered as an important systematic noise floor in the measurement of the absolute, white-light transit depth. We do not detect this variability in the individually measured transit depths (Tab.~\ref{tab:visits}) because it is smaller than the uncertainty on each. Most importantly, while the spot-induced variability influences the absolute depth at each epoch, its effect on the relative transit depth among wavelengths will be much smaller and not substantially bias our transmission spectrum estimate.

\subsubsection{Modeling Instrumental Systematics}\label{sec:ramp}
Before calculating GJ1214b's transmission spectrum, we detour slightly to use the white light curves's high photometric precision to investigate the characteristics of WFC3's instrumental systematics. Rather than correcting for the instrumental systematics with the simple non-parametric \divideoot~method, in this section we describe them with an analytic model whose parameters illuminate the physical processes at play. We refer to this treatment as the \modelramp~method. 

In this model, we treat the systematics as consisting of an exponential ramp, an orbit-long slope, and a visit-long slope. We relate the observed flux ($F_{\rm obs}$) to the systematics-free flux ($F_{\rm cor}$) by
\begin{equation}
\frac{F_{\rm obs}}{F_{\rm cor}} = (C + Vt_{\rm vis} + Bt_{\rm orb})\left(1 - Re^{-(t_{\rm bat} - D_{\rm b})/\tau}\right)
\end{equation}
where $t_{\rm vis}$ is time within a visit ($=0$ at the middle of each visit), $t_{\rm orb}$ is time within an orbit ($=0$ at the middle of each orbit),  $t_{\rm bat}$ is time within a batch ($=0$ at the start of each batch), \edit{$\tau$ is a ramp timescale}, and the term
\begin{equation}
 D_{\rm b} = \left\{ 
  \begin{array}{l l}
    D & \quad \text{for the \first~batch of an orbit}\\
    0 & \quad \text{for the other batches}\\
  \end{array} \right.
  \label{eq:ramp}
\end{equation}
allows the exponential ramp to be delayed slightly for the first batch of an orbit.

The exponential form arises out of the toy model proposed by \citet{agol.2010.c1fftems}, where a certain volume of the detector pixels has the ability to temporarily trap charge carriers and later release them as excess dark current. In quick series of sufficiently strong exposures, the population of charge traps approaches steady state, corresponding to the flattening of the exponential. Judging by the appearance of the ramp in the \second-\fourth~batches of each orbit, the release timescale seems to be short enough that the trap population completely resets to the same baseline level after each 6 minute buffer download (during which the detector was being continually flushed each 2.9 seconds). Compared to these batches, the \first~batch of each orbit appears to exhibit a ramp that is either weaker, or as we have parameterized it with the $D_b$ term, delayed. We do not explain this, but we hypothesize that it relates to rapid changes in the physical state of the detector coming out of Earth occultation affecting the pixels' equilibrium charge trap populations.

The visit-long and orbit-long slopes are purely descriptive terms \citep[as in][]{brown.2001.hsttptp2,carter.2009.ntpe1,nutzman.2011.peppes1ehstfgstao}, but relate to physical processes in the telescope and camera. The orbit-long slope probably arises from the combination of pointing/focus drifts (see Fig.~\ref{fig:timeseries}) with our imperfect flat-fielding of the detector. This effect of this orbital phase term could be equally well-achieved, for instance, by including a linear function of the \zeroth~order $x$ and $y$ positions \citep[see][]{burke.2010.notjx,pont.2007.hsttppt1mrs,swain.2008.pmaep}. The visit-long slope is not mirrored in any of the measured geometrical properties of the star on the detector, and is more difficult to associate with a known physical cause.

In order to determine the parameters $C$, $V$, $B$, $R$, $D$, and $\tau$, we fit Eq.~\ref{eq:ramp} multiplied by a transit model to the last three orbits of each visit's uncorrected white light curve. The transit parameters are allowed to vary exactly as in \S\ref{sec:individual}, including the use of the informative prior on the limb-darkening coefficients. The white light curves with the best \modelramp~fit are shown in Fig.~\ref{fig:white}, and the \edit{properties of the} residuals from this model are shown in Fig.~\ref{fig:residuals}. The transit parameters from this independent systematics correction method are consistent with those in Tab~\ref{tab:visits}.  We do not quite achieve the 280 ppm predicted scatter in the \modelramp~light curves, and the residuals show slight evidence for correlated noise (Fig.~\ref{fig:residuals}). More complicated instrumental correction models could almost certainly improve this, but we only present this simple model for heuristic purposes. In all sections except this one, we use the \divideoot-corrected data exclusively for drawing scientific conclusions about GJ1214b. 

\begin{figure}[t] 
   \centering
   \includegraphics[width=\columnwidth]{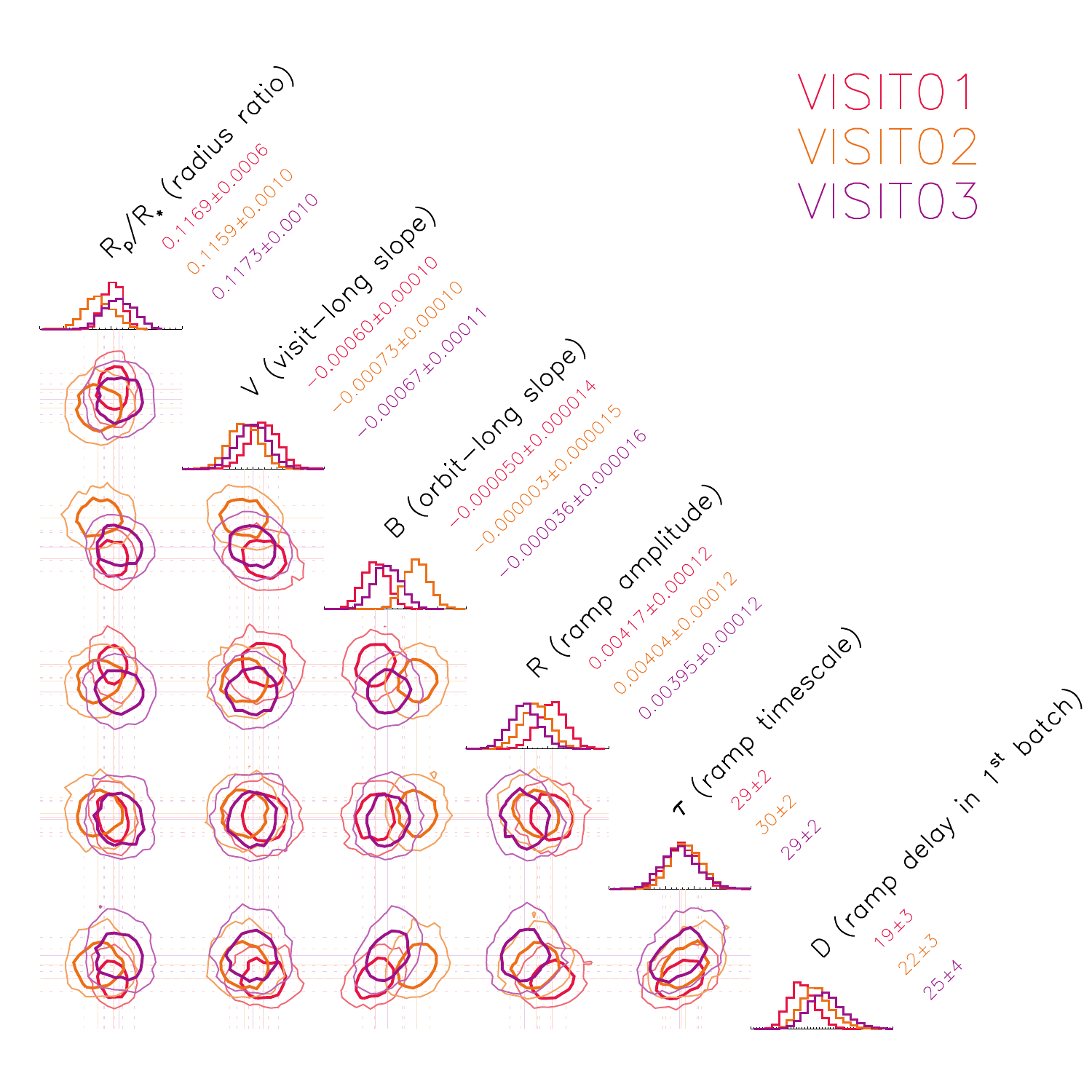} 
   \caption{The {\it a posteriori} distribution of the instrumental systematics parameters from the analytic model, in each of the three visits. The MCMC results for single parameters ({\em diagonal;} histograms) and pairs of parameters ({\em off-diagonal;} contours encompassing 68\% and 95\% of the distribution) are shown, marginalized over all other parameters (including $c$ and $d$ with priors, $t_{14}$, and $R_\star$).  $V$ is measured in units of relative flux/$(3\times96~{\rm minutes})$, $B$ in relative flux/$(96~{\rm minutes})$, $R$ in relative flux, and both $\tau$ and $D$ in seconds. All visits are plotted on the same scale; for quantitative comparison, the median values and $1\sigma$ uncertainties of each parameter are quoted along the diagonal. The systematics parameters are remarkably repeatable from visit to visit; also, they are largely uncorrelated with \rp/\rs~({\em left column}). }
   
   \label{fig:systematics}
   \end{figure}

Fig.~\ref{fig:systematics} shows the inferred PDF's of the instrumental systematics parameters for all three HST visits, graphically demonstrating the striking repeatability of the systematics. As expected from the nearly identical cadence of illumination within each of the three visits, the ramp has the same $R=0.4\%$ amplitude, $\tau=30$ second timescale, and $D=20$ second delay time across all observations. The values of $\tau$ and $D$ are similar to the time for a single exposure, 25 seconds (including overhead). While the visit-long slope $V$ is of an amplitude (fading by 0.06\% over an entire visit) that could conceivably be consistent with stellar variability, the fact that it is identical across all three visits argues strongly in favor of it being an instrumental systematic. $B$ is the only parameter that shows any evidence for variability between orbits; we would expect this to be the case if this term arises out of flat-fielding errors, since the \first~order spectrum falls on different pixels in the three visits.

\subsection{Spectroscopic Light Curve Fits}\label{sec:scanalysis}
We construct multiwavelength spectroscopic light curves by binning the extracted first order spectra into channels that are 5 pixels ($\Delta \lambda = 23$ nm) wide. We estimate the flux, flux uncertainty, and effective wavelength of each bin from the inverse-variance (estimated from the noise model) weighted average of each quantity over the binned pixels. For each of these binned spectroscopic light curves, we employ the \divideoot~method to correct for the instrument systematics. 

To measure the transmission spectrum of GJ1214b, we fit each of these 24 spectroscopic light curves from each of the three visits with a model in which \rp/\rs, $c$, $d$, and $s$ are allowed to vary. We hold the remaining parameters fixed so that $a/R_\star = 14.9749 $ and $b = 0.27729$, which are the values used by \citet{bean.2010.gtsse1}, \citet{desert.2011.oemasg}, and \citet{croll.2011.btss1smmwa}. For limb-darkening priors, we use the same sized Gaussians on the same linear combinations of $c$ and $d$ as in \S\ref{sec:combined}, but center them on the {\tt PHOENIX}-determined best values for each spectroscopic bin (see \S\ref{sec:limbdarkening}). The correlation length of all parameters is $<10$ in the MCMC chains.

For most spectroscopic bins, the inferred value of $s$ is within $1\sigma$ of unity, indicating that the flux residuals show scatter commensurate with that predicted from photon noise (1400 to 1900 ppm across wavelengths). No evidence for correlated noise is seen in any of the bins, as judged by the same criterion as for the white light curves (see Fig.~\ref{fig:residuals}). 

\begin{figure*}[t] 
   \centering
   \includegraphics[width=\textwidth]{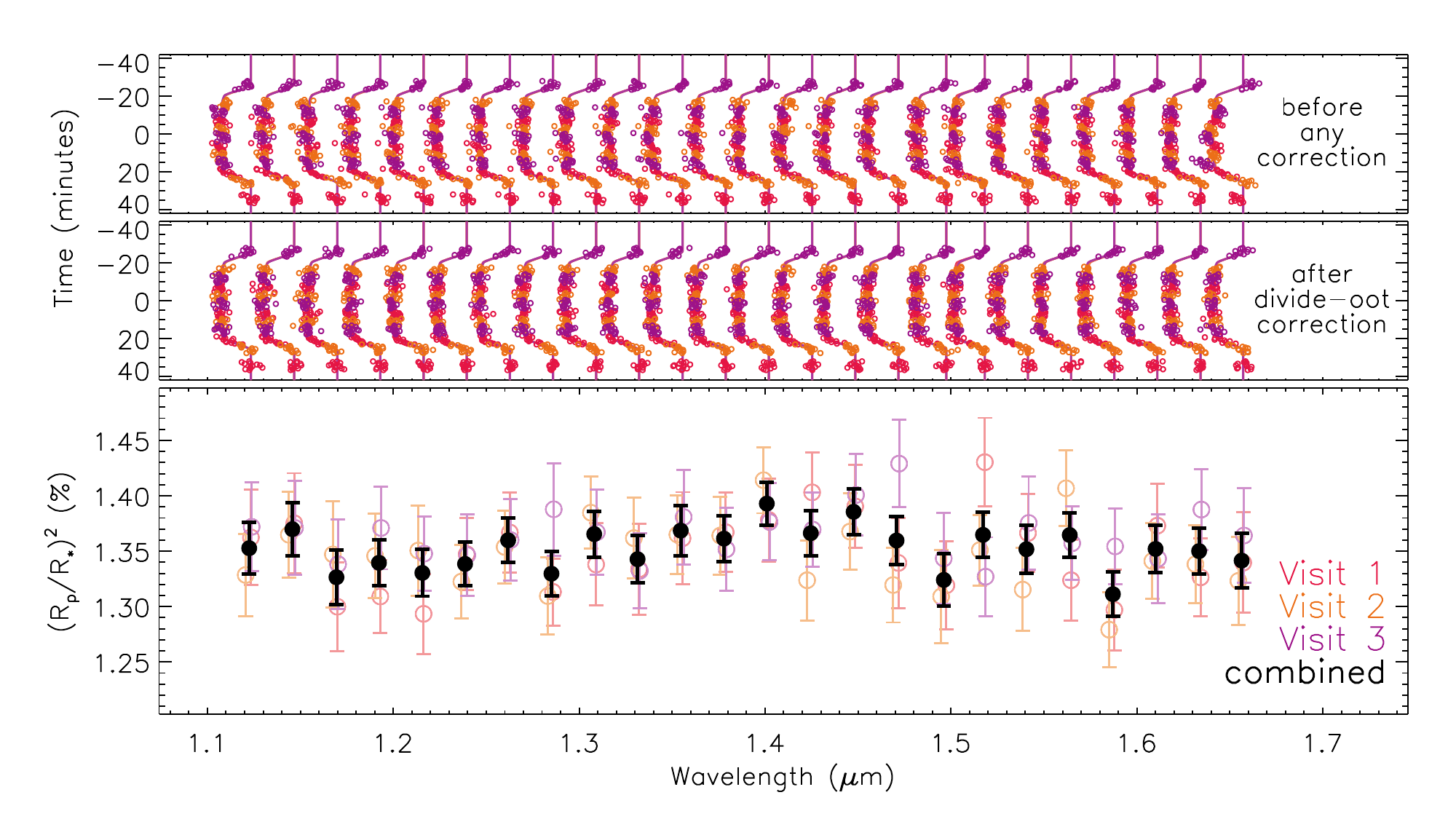} 
   \caption{{\em Top panels:} Spectroscopic transit light curves for GJ1214b, before and after the \divideoot~correction, rotated and offset for clarity. {\em Bottom panel:} The combined transmission spectrum of GJ1214b ({\em black circles with error bars}), along with the spectra measured for each visit ({\em colorful circles}). Each light curve in the top panel is aligned to its respective wavelength bin in the panel below. Colors denote HST visit throughout.}
   \label{fig:transmissionspectrum}
\end{figure*}

Fig.~\ref{fig:transmissionspectrum} shows the transmission spectra inferred from each of the three visits, as well as the \divideoot-corrected, spectrophotometric light curves from which they were derived. For the final transmission spectrum (shown as black points in Fig.~\ref{fig:transmissionspectrum}), we combine the three values of \rp/\rs, and $\sigma_{R_p/R_\star}$ in each wavelength bin by averaging them over the visits with a weighting proportional to $1/\sigma_{R_p/R_\star}^2$. Table~\ref{tab:spectrum} gives this average transmission spectrum, as well as the central values of the limb-darkening prior used in each bin. The wavelength grids in the three visits are offset slightly (by less than a pixel) from one another; in Table~\ref{tab:spectrum} we quote the average wavelength for each bin. 

In \S\ref{sec:individual} we found that GJ1214's 1\% variability at WFC3 wavelengths causes $\Delta D = 0.014\%$ or $\Delta R_p/R_\star = 0.0006$ variations in the absolute transit depth. The starspots causing this variability would have a similar effect on measurements of the transmission spectrum, but unless GJ1214's starspot spectrum is maliciously behaved, the offsets should be broad-band and  the influence on the wavelength-to-wavelength variations within the WFC3 transmission spectrum should be much smaller. Each visit's transmission spectrum is a differential measurement made with respect to the integrated stellar spectrum at each epoch; by averaging together three estimates to produce our final transmission spectrum, we average over the time-variable influence of the starspots. Importantly, if GJ1214 is host to a large population of starspots that are symmetrically distributed around the star and do not appear contribute to the observed flux variability over the stellar rotation period, their effect on the transmission spectrum will not average out \citep[see][]{desert.2011.tse1isom, carter.2011.tlcpxsts1, berta.2011.gsssvtsap}.

If we fix the limb-darkening coefficients to the {\tt PHOENIX} values instead of using the prior, the uncertainties on the \rp/\rs~measurements decrease by 20\%. If we use only a single pair of LD coefficients (those for the white light curve) instead of those matched to the individual wavelength bins, the transmission spectrum changes by about $1\sigma$ \edit{on the individual bins}, in the direction of showing stronger water features and being less consistent with an achromatic transit depth. These tests confirm that the presence of the broad \hho~feature in the stellar spectrum (see Fig.~\ref{fig:spectrum}) makes it especially {\em crucial} that we employ the detailed, multiwavelength LD treatment. 

As a test to probe the influence of the \divideoot~systematics correction, we repeat this section's analysis using the analytic \modelramp~method to remove the instrumental systematics; every point in the transmission spectrum changes by much less than $1\sigma$. \edit{We also experimented with combining the three visits' spectroscopic light curves and fitting for them jointly, instead of averaging together the transmission spectra inferred separately from each visit. We found the results to be practically identical to those quoted here.}

Because the transmission spectrum is conditional on the orbital parameters we held fixed ($a/R_\star$, $b$), we underestimate the uncertainty in the {\em absolute} values of \rp/\rs~; the quoted $\sigma_{R_p/R_\star}$ are intended for {\em relative} comparisons only. Judging by the $R_p/R_\star$ uncertainty in the unconstrained white light curve fit (Table~\ref{tab:combinedfit}), varying $a/R_\star$ could cause the ensemble of $R_p/R_\star$ measurements in Table~\ref{tab:spectrum} to move up or down in tandem with a systematic uncertainty that is comparable to the statistical uncertainty on each. This is in addition to the $\Delta R_p/R_\star = 0.0006$ offsets expected from stellar variability \citep{berta.2011.gsssvtsap}.

\subsection{Searching for Transiting Moons}\label{sec:moons}

Finally, we search for evidence of transiting satellite compansions to GJ1214b in our summed WFC3 light curves. The light curve morphology of transiting exomoons can be complicated, but they could generally appear in our data as shallow transit-shaped dimmings {\em or} brightenings offset from the planet's transit light curve \citep[see][for a detailed discussion]{kipping.2011.lagdpt}. While the presence of a moon could also be detected in temporal variations of the planetary transit duration \citep{kipping.2009.ttee}, we only poorly constrain GJ1214b's transit duration in individual visits due to incomplete coverage.

Based on the Hill stability criterion, we would not expect moons to survive farther than 8 planetary radii away from GJ1214b so their transits should not be offset from GJ1214b's by more than 25 minutes, less than the duration of an HST Earth occultation. We search only the data in the in-transit visit, using the \divideoot~method to correct for the systematics. Owing to the long buffer download gaps in our light curves (see \S\ref{sec:observations}), the most likely indication of a transiting moon in the WFC3 light curve would be an offset in flux from one 12-exposure batch to another. Given the \edit{376 ppm} per-exposure scatter in the \divideoot~corrected light curve, we would have expected to be able to identify transits of 0.4 \rearth~(Ganymede-sized) moons at $3\sigma$ confidence. We see no strong evidence for such an offset. Also, we note that starspot occultations could easily mimic the  light curve of a transiting exomoon in the time coverage we achieve with WFC3, and such occultations are known to occur in the GJ1214b system \citep[see][]{berta.2011.gsssvtsap,carter.2011.tlcpxsts1,kundurthy.2011.ao1spesa}. 

Due to the many possible configurations of transiting exomoons and the large gaps in our WFC3 light curve, our non-detection of moons does not by itself place strict limits on the presence of exo-moons around GJ1214b. 

\section{Discussion} \label{sec:discussion}

The average transmission spectrum of GJ1214b from our three HST visits is shown in Fig.~\ref{fig:transmissionspectrum}. To the precision afforded by the data, this transmission spectrum is flat; a simple weighted mean of the spectrum is a good fit, with $\chi^2=20.4$ for 23 degrees of freedom.

 \subsection{Implications for Atmospheric Compositions}
 
We compare the WFC3 transmission spectrum to a suite of cloud-free theoretical atmosphere models for GJ1214b. The models were calculated in \citet{miller-ricci.2010.nats1}, and we refer the reader to that paper for their details. To compare them to our transmission spectrum, we bin these high-resolution ($R=1000$) models to the effective wavelengths of the 5-pixel WFC3 spectroscopic channels ($R = 50-70$) by integrating over each bin. Generally, to account for the possible suppression of transmission spectrum features caused by the overlap of shared planetary and stellar absorption lines, this binning should be weighted by the photons detected from the system at very high resolution, but this added complexity is not justified for our dataset. The normalization of the model spectra is uncertain  (i.e. the planet's true \rp), so we allow a multiplicative factor in \rp/\rs~to be applied to each (giving 24-1=23 degrees of freedom for all models). Varying the bin size between 2 and 50 pixels wide does not significantly change any of the results we quote in this section.

A solar composition atmosphere in thermochemical equilibrium is a terrible fit to the WFC3 spectrum; it has a \chisq=126.2 (see Fig.~\ref{fig:transmissionspectrum}) and is formally ruled out at 8.2$\sigma$ confidence. Likewise, the same atmosphere but enhanced $50\times$ in elements heavier than helium, a qualitative approximation to the metal enhancement in the Solar System ice giants \citep[enhanced $30-50\times$ in C/H;][]{gautier.1995.tn,encrenaz.2005.nagpocm, guillot.2009.gp}, is ruled out at $7.5\sigma$ ($\chi^2 = 113.2$). Both models assume equilibrium molecular abundances and the absence of high-altitude clouds; if GJ1214b has an \hh-rich atmosphere, at least one of these assumptions would have to be broken. 

\begin{figure*}[t] 
   \centering
   \includegraphics[width=\textwidth]{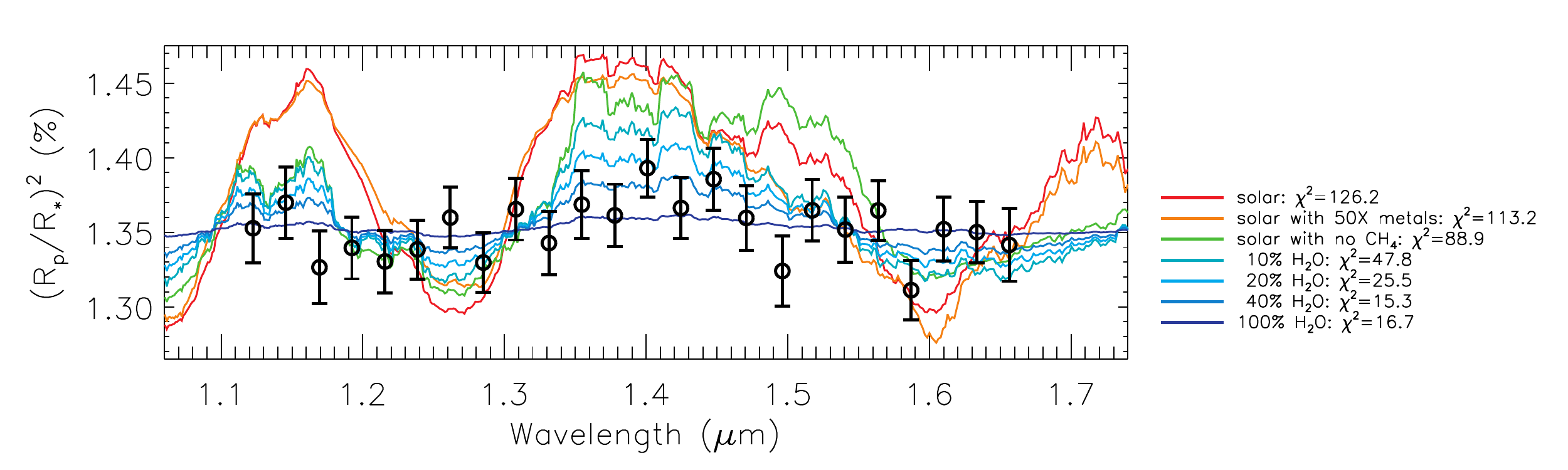} 
   \caption{The WFC3 transmission spectrum of GJ1214b ({\em black circles with error bars}) compared to theoretical models ({\em colorful lines}) with a variety of compositions. The high resolution models are shown here smoothed for clarity, but were binned over each measured spectroscopic bin for the \chisq~comparisons. The amplitude of features in the model transmission spectra increases as the mean molecular weight decreases between a 100\% water atmosphere ($\mu=18$) and a solar composition atmosphere ($\mu=2.36$).}
   \label{fig:wfc3_models}
\end{figure*}

Suggesting, along these lines, that photochemistry might deplete GJ1214b's atmosphere of methane, \citet{desert.2011.oemasg}, \citet{croll.2011.btss1smmwa}, and \citet{crossfield.2011.hdnts1} have noted their observations to be consistent with a solar composition model in which \chhhh~has been artificially removed. With the WFC3 spectrum alone, we can rule out such an \hh-rich, \chhhh-free atmosphere at 6.1$\sigma$ (Fig.~\ref{fig:wfc3_models}). This is consistent with \citet{miller-ricci-kempton.2011.ac1pc}'s theoretical finding that such thorough methane depletion cannot be achieved through photochemical processes, even when making extreme assumptions for the photoionizing UV flux from the star.

\begin{figure}[t] 
   \centering
   \includegraphics[width=\columnwidth]{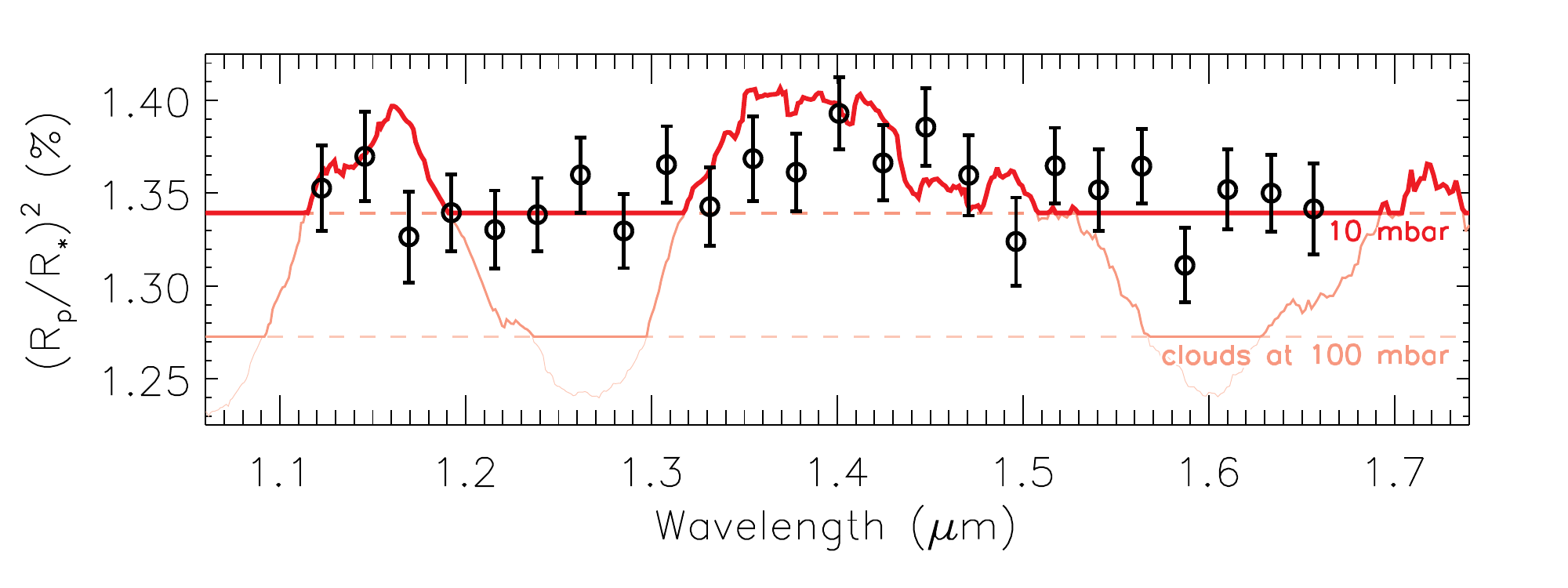} 
   \caption{The WFC3 transmission spectrum of GJ1214b ({\em black circles with error bars}) compared to a model solar composition atmosphere that has thick clouds located at altitudes of 100 mbar ({\em pink lines}) and 10 mbar {\em red lines}). We treat the hypothetical clouds in an {\em ad hoc} fashion, simply cutting off transmission through that atmosphere below the denoted pressures. }
   \label{fig:wfc3_clouds}
\end{figure}

Previous spectroscopic measurements in the red optical \citep{bean.2010.gtsse1,bean.2011.ontsspgfema} could only be reconciled with a \hh-rich atmosphere if such an atmosphere were to host a substantial cloud layer at an altitude above 200 mbar \citep[see][]{miller-ricci-kempton.2011.ac1pc}. How far the flattening influence of such a cloud layer would extend beyond 1 \micron~to WFC3 wavelengths would depend on both the concentration and size distribution of the scattering particles. \edit{As such, we explore} possible cloud scenarios consistent with the WFC3 spectrum in an {\em ad hoc} fashion, using a solar composition atmosphere and arbitrarily cutting off transmission below various pressures to emulate optically thick cloud decks at different altitudes in the atmosphere. Fig.~\ref{fig:wfc3_clouds} summarizes the results. A cloud deck at 100 mbar, which would be sufficient to flatten the red optical spectrum, is ruled out at $5.7 \sigma$ ($\chi^2 = 82.8)$. \edit{Due to higher opacities between 1.1 and 1.7 \micron,} WFC3 probes higher altitudes in the atmosphere than the red optical, requiring clouds closer to 10 mbar to match the data ($\chi^2=23.4$).  Note, with the term ``clouds'' we refer to all types \edit{of particles that cause broad-band extinction, whether they scatter or absorb,} and whether they were formed through near-equilibrium condensation (such as Earth's water clouds) or through upper atmosphere photochemistry (such as Titan's haze). 

 \citet{fortney.2005.ecctpawts} and \citet{miller-ricci-kempton.2011.ac1pc} identified KCl and ZnS as condensates that would be likely to form in GJ1214b's atmosphere, but found they would condense deeper in the atmosphere (200-500 mbar) than required by the WFC3 spectrum \edit{and would probably not be optically thick}. While winds may be able to loft such clouds to higher altitudes, it is not clear that the abundance of these species alone would be sufficient to blanket the entire limb of the planet with optically thick clouds. The condensation and complicated evolution of clouds has been studied within the context of cool stars and hot Jupiters \citep[e.g][]{lodders.2006.cmso,helling.2008.dbdepiccsaqcl}, but further study into the theoretical landscape for equilibrium clouds on planets in GJ1214b's \edit{gravity} and temperature regime is certainly warranted. The scattering may also be due to a high altitude haze formed as by-products of high-altitude photochemistry; \citet{miller-ricci-kempton.2011.ac1pc} found the conditions on GJ1214b to allow for the formation of complex hydrocarbon clouds through methane photolysis.

However such clouds might form, they would either need to be optically thick up to a well-defined altitude or consist of a substantial distribution of particles \edit{acting in the Mie regime, i.e. with sizes approaching 1 \micron. Neither the VLT spectra nor our observations give any definitive indications of the smooth falloff in transit depth toward longer wavelengths that would be expected from Rayleigh scattering by molecules or small particles. This is unlike the case of the hot Jupiter HD189733b, where the uniform decrease in transit depth from 0.3 to 1 \micron~\citep{pont.2008.dahep0mts1wh,lecavelier-des-etangs.2008.rsts1,sing.2011.hsttse1hahonws}} and perhaps to as far as 3.6 \micron~\citep[see][]{sing.2009.tse1iswfhwn, desert.2009.scmate1} has been convincingly attributed to a small particle haze.

As an alternative, the transmission spectrum of GJ1214b could be flat simply because the atmosphere has a large mean molecular weight. We test this possibility with \hh~atmospheres that contain increasing fractions of \hho. This is a toy model, but including  molecules other than \hh~or \hho~in the atmosphere would serve principally to increase $\mu$ without substantially altering the opacity between 1.1 and 1.7 \micron, so the limits we place on $\mu$ are robust. We find that an atmosphere with a 10\% water by number (50\% by mass) is disfavored by the WFC3 spectrum at $3.1\sigma$ ($\chi^2 = 47.8$), as shown in Fig.~\ref{fig:transmissionspectrum}. All fractions of water above 20\% (70\% by mass) are good fits to the data ($\chi^2 < 25.5$). The $10\%$ water atmosphere would have a minimum mean molecular weight of $\mu  = 3.6$, which we take as a lower limit on the atmosphere's mean molecular weight. 

\begin{figure*}[t] 
   \centering
   \includegraphics[width=\textwidth]{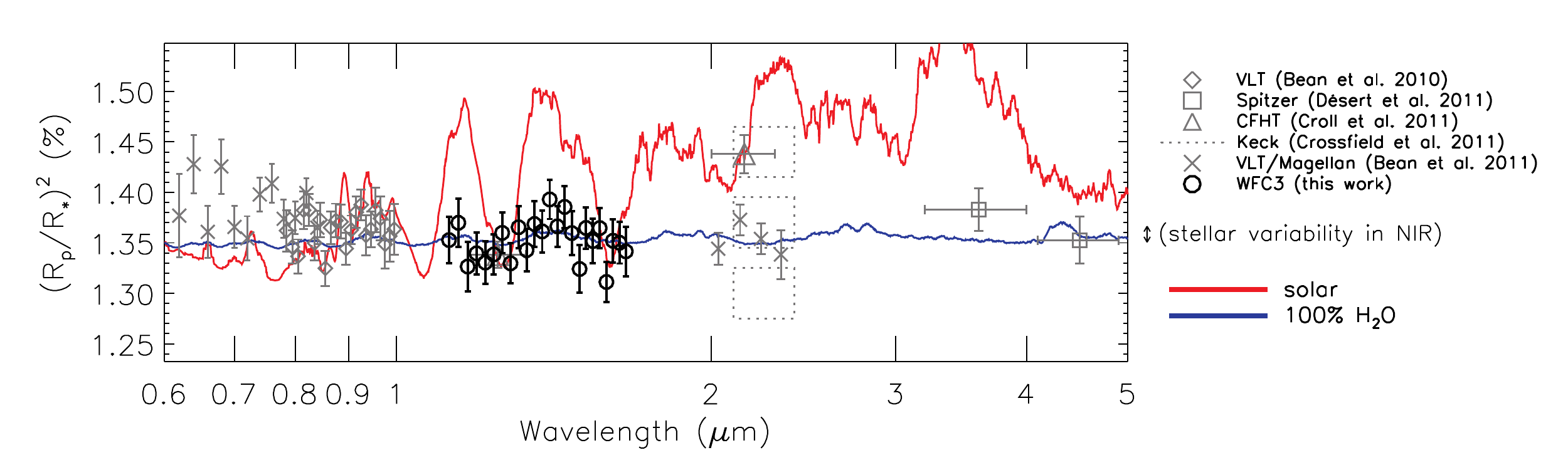} 
   \caption{GJ1214b's transmission spectrum from WFC3 in the context of observations from the VLT \citep[0.6-1 \micron;][]{bean.2010.gtsse1, bean.2011.ontsspgfema}, CFHT \citep[1.25 + 2.15 \micron;][]{croll.2011.btss1smmwa}, Magellan \citep[2.0-2.3 \micron][]{bean.2011.ontsspgfema}, and Spitzer \citep[3.6 + 4.5 \micron;][]{desert.2011.oemasg}. While they do not measure an absolute transit depth, observations from NIRSPEC on Keck \citep[2.1-2.3 \micron][]{crossfield.2011.hdnts1} disfavor models they tested that had amplitudes larger than 0.05\% in their wavelength range; we represent these constraints with the dashed rectangular boxes. Two extremes of the models explored in this paper are shown, normalized to the MEarth-measured transit depth \citep[see][]{miller-ricci.2010.nats1}. It is important to note that stellar variability could cause individual data sets to shift up or down on this plot as much as $\Delta D = 0.014\%$ in the near-IR or $2-3\times$ more in the optical, depending on the stellar spot spectrum. }
\label{fig:combinedtransmissionspectrum}
\end{figure*}

For the sake of placing the WFC3 transmission spectrum in the context of other observations of GJ1214b, we also display it alongside the published transmission spectra from the VLT \citep{bean.2010.gtsse1,bean.2011.ontsspgfema}, CFHT \citep{croll.2011.btss1smmwa}, Magellan \citep{bean.2011.ontsspgfema}, and Spitzer \citep{desert.2011.oemasg} in Fig.~\ref{fig:combinedtransmissionspectrum}. Stellar variability could cause individual sets of observations to move up and down on this plot by as much as $\Delta D = 0.014\%$ for measurements in the near-IR \citep{berta.2011.gsssvtsap}; we indicate this range of potential offsets by an arrow at the right of the plot. We display the measurements in Fig.~\ref{fig:combinedtransmissionspectrum} with no relative offsets applied and note that their general agreement is consistent with the predicted small influence of stellar variability. Depending on the temperature contrast of the spots, however, the variability could be larger by a factor of $2-3\times$ in the optical, and we caution the reader to consider this systematic uncertainty when comparing depths between individual studies. \edit{For instance, the slight apparent rise in \rp/\rs~toward 0.6 \micron, that would potentially be consistent with Rayleigh scattering in a low-$\mu$ atmosphere, could also be easily explained through the poorly constrained behavior of the star in the optical. Indeed, \citet{bean.2011.ontsspgfema} found a significant offset between datasets that overlap in wavelength (near 0.8 \micron) but were taken in different years, suggesting variability plays a non-negligible role at these wavelengths.}
 
\edit{Finally, we note that any model with $\mu > 4$, such as one with a $> 50\%$ mass fraction of water},  would be consistent with the measurements from \citet{bean.2010.gtsse1}, \citet{desert.2011.oemasg}, \citet{crossfield.2011.hdnts1}, \citet{bean.2011.ontsspgfema}, and WFC3. The only observation it could not explain would be the deep $K_s$-measurement from \citet{croll.2011.btss1smmwa}. Of the theoretical models we tested, we could find none that matched all the available measurements. We are uncertain of how to interpret this apparent incompatibility but hopeful that future observational and theoretical studies of the GJ1214b system may clarify the issue. In the meantime, we adopt an atmosphere with at least 50\% water by mass as the most plausible model to explain the WFC3 observations. 

\subsection{Implications for GJ1214b's Internal Structure}

If GJ1214b is not shrouded in achromatically optically thick high-altitude clouds, the WFC3 transmission spectrum disfavors any proposed bulk composition for the planet that relies on a substantial, unenriched, hydrogen envelope to explain the planet's large radius. Both the ice-rock core with nebular H/He envelope and pure rock core with outgassed \hh~envelope scenarios explored by \citet{rogers.2010.tpol1} would fall into this category, requiring additional ingredients to match the observations. In contrast, their model that achieves GJ1214b's large radius mostly from a large water-rich core, would agree with our observations.

Perhaps most compellingly, a high $\mu$ scenario would be consistent with composition proposed by \citet{nettelmann.2011.tesmts1}, who found that GJ1214b's radius could be explained by a bulk composition consisting of an ice-rock core surrounded by a H/He/\hho~envelope that has a water mass fraction of 50-85\%. Such a composition would be intermediate between the H/He- and \hho-envelope limiting cases proposed by \citet{rogers.2010.tpol1}. The H/He/\hho~envelope might arise if GJ1214b had originally accreted a substantial mass of hydrogen and helium from the primordial nebula but then was depleted of its lightest molecules through atmospheric escape.

\setlength{\tabcolsep}{0.04in} 
\begin{deluxetable}{cccc}
\tablecaption{\label{tab:spectrum}GJ1214b's Transmission Spectrum from WFC3/G141}
\tabletypesize{\normalsize}
\tablecolumns{4}
\tablehead{\colhead{Wavelength (\micron)} & \colhead{$R_p/R_\star$} & \colhead{$c$} & \colhead{$d$}}
\startdata
$1.123  $ &   $0.11641\pm0.00102 $  &$  -0.372 $  &$   1.068 $  \\
$1.146  $ &   $0.11707\pm0.00099 $  &$  -0.397 $  &$   1.088 $  \\
$1.169  $ &   $0.11526\pm0.00098 $  &$  -0.404 $  &$   1.089 $  \\
$1.192  $ &   $0.11589\pm0.00093 $  &$  -0.410 $  &$   1.090 $  \\
$1.215  $ &   $0.11537\pm0.00091 $  &$  -0.406 $  &$   1.075 $  \\
$1.239  $ &   $0.11574\pm0.00090 $  &$  -0.403 $  &$   1.063 $  \\
$1.262  $ &   $0.11662\pm0.00088 $  &$  -0.407 $  &$   1.064 $  \\
$1.285  $ &   $0.11565\pm0.00088 $  &$  -0.403 $  &$   1.045 $  \\
$1.308  $ &   $0.11674\pm0.00085 $  &$  -0.411 $  &$   1.042 $  \\
$1.331  $ &   $0.11595\pm0.00087 $  &$  -0.390 $  &$   1.068 $  \\
$1.355  $ &   $0.11705\pm0.00089 $  &$  -0.368 $  &$   1.101 $  \\
$1.378  $ &   $0.11664\pm0.00088 $  &$  -0.371 $  &$   1.110 $  \\
$1.401  $ &   $0.11778\pm0.00088 $  &$  -0.338 $  &$   1.075 $  \\
$1.425  $ &   $0.11693\pm0.00091 $  &$  -0.295 $  &$   1.029 $  \\
$1.448  $ &   $0.11772\pm0.00090 $  &$  -0.319 $  &$   1.056 $  \\
$1.471  $ &   $0.11663\pm0.00092 $  &$  -0.322 $  &$   1.061 $  \\
$1.496  $ &   $0.11509\pm0.00100 $  &$  -0.345 $  &$   1.084 $  \\
$1.517  $ &   $0.11635\pm0.00104 $  &$  -0.305 $  &$   1.013 $  \\
$1.541  $ &   $0.11626\pm0.00091 $  &$  -0.330 $  &$   1.042 $  \\
$1.564  $ &   $0.11681\pm0.00091 $  &$  -0.351 $  &$   1.059 $  \\
$1.587  $ &   $0.11443\pm0.00091 $  &$  -0.372 $  &$   1.072 $  \\
$1.610  $ &   $0.11631\pm0.00091 $  &$  -0.399 $  &$   1.082 $  \\
$1.633  $ &   $0.11620\pm0.00092 $  &$  -0.399 $  &$   1.075 $  \\
$1.656  $ &   $0.11581\pm0.00096 $  &$  -0.415 $  &$   1.081 $  \\
\enddata
\end{deluxetable}

\subsection{Prospects for GJ1214b}\label{sec:future}
Future observations with the James Webb Space Telescope \edit{\citep{deming.2009.dctseuatsfjwst, kaltenegger.2009.tep}}, one of the immense next generation ground-based telescopes \edit{\citep[GMT, TMT, ELT; see][]{ehrenreich.2006.tsetp}}, or possibly even a dedicated campaign with current facilities, could detect the 0.01\% transmission spectrum features of a 100\% water atmosphere on GJ1214b, and potentially distinguish between clear \hh-poor and cloudy \hh-rich atmospheres. Along another front, simulations by \citet{menou.2011.accg} show that observations of GJ1214b's thermal phase curve, such as those for HD189733b by \citet{knutson.2007.dcep1}, would probe the ratio of radiative to advective timescales in GJ1214b's outer envelope and provide an independent constraint on the atmospheric composition. Detecting the thermal emission from this 500K exoplanet is currently very difficult, and will likely have to wait until the launch of JWST.

In the meantime, we advocate further study of the GJ1214 system in general. Confirming and refining the parallax for the system \citep{van-altena.1995.gctsp} will improve our knowledge of the stellar mass, and in turn, the planet's mass and radius. Likewise, further radial velocity observations will empirically constrain the hypothesis by \citet{carter.2011.tlcpxsts1} that a significantly non-zero orbital eccentricity could be biasing GJ1214b's inferred density.

\section{Conclusions} \label{sec:conclusions}
In this work, we made new measurements of the GJ1214b's transmission spectrum using HST/WFC3. Reaching a precision of $\sigma_{R_p/R_\star}=0.0009$ in 24 simultaneously measured wavelength bins, we found the transmission spectrum to be completely flat between 1.1 and 1.7 \micron. We saw no evidence for the strong \hho~absorption features expected from a range of \hh-rich model atmospheres. 

Given the lack of a known source for clouds or hazes that could create a truly achromatic transit depth across all wavelengths, we interpret this flat WFC3 transmission spectrum to be best explained by an atmosphere with a high mean molecular weight.  Based on our observations, this atmosphere would likely consist of more than 50\% water by mass or a mean molecular weight of $\mu > 4$. Such an atmosphere would be consistent with observations of GJ1214b's transmission spectrum by \citet{bean.2010.gtsse1}, \citet{desert.2011.oemasg}, \citet{crossfield.2011.hdnts1}, and \citet{bean.2011.ontsspgfema} although it would be difficult to reconcile with those by \citet{croll.2011.btss1smmwa}. 

Such a constraint on GJ1214b's upper atmosphere serves as a boundary condition for models of bulk composition and structure of the rest of the planet. It suggests GJ1214b contains a substantial fraction of water throughout the interior of the planet in order to obviate the need for a completely H/He- or \hh-dominated envelope to explain the planet's large radius. A high bulk volatile content would point to GJ1214b forming beyond the snow line and migrating inward, although any such statements about GJ1214b's past are subject to large uncertainties in the atmospheric mass loss history \citep[see][]{rogers.2011.fsle}.

Finally, this paper is the first published study using WFC3 for observing a transiting exoplanet. Aside from several instrumental systematics that were straightforward to correct and did not require a detailed instrumental model, the camera delivered nearly photon-limited performance both in individual spectrophotometric light curves and in summed white light curves. We are confident that WFC3 will serve as a valuable tool for exoplanet atmospheric characterization in the years to come.

\acknowledgements

We thank Jacob Bean, Bryce Croll, Ian Crossfield, Ernst de Mooij, Leslie Rogers, Drake Deming, Avi Mandell, Martin K\"ummel, Ron Gilliand, Tom Wheeler, and the CfA Summer Statistics Club for valuable discussions regarding this work. We are extremely grateful to our Program Coordinator Patricia Royle, Contact Scientist Howard Bushhouse, the WFC3 instrument team, the entire staff at STScI and NASA, and the crew of STS-125 for their crucial roles in enabling these observations. \edit{We thank the anonymous referee for a thorough reading and thoughtful comments that improved the manuscript.} E.K. acknowledges funding from NASA through the Sagan Fellowship Program.  P.R.M. thanks CfA director Charles Alcock for enabling a sabbatical at SAO. We gratefully acknowledge funding from the David and Lucile Packard Fellowship for Science and Engineering (awarded to D.C.), the National Science Foundation (grant  AST-0807690, awarded to D.C.), and NASA (grant HST-GO-12251). This work is based on observations made with the NASA/ESA Hubble Space Telescope, obtained at the Space Telescope Science Institute, which is operated by the Association of Universities for Research in Astronomy, Inc., under NASA contract NAS 5-26555. These observations are associated with program \#GO-12251. This research has made use of NASA's Astrophysics Data System.

{\it Facilities:} \facility{HST (WFC3)}


\begin{thebibliography}{90}
\expandafter\ifx\csname natexlab\endcsname\relax\def\natexlab#1{#1}\fi


\bibitem[{{Agol} {et~al.}(2010){Agol}, {Cowan}, {Knutson}, {Deming}, {Steffen},
  {Henry}, \& {Charbonneau}}]{agol.2010.c1fftems}
{Agol}, E., {Cowan}, N.~B., {Knutson}, H.~A.,  {et~al.} 2010, \apj, 721, 1861

\bibitem[{{Batalha} {et~al.}(2011){Batalha}, {Borucki}, {Bryson}, {Buchhave},
  {Caldwell}, {Christensen-Dalsgaard}, {Ciardi}, {Dunham}, {Fressin},
  {Gautier}, {Gilliland}, {Haas}, {Howell}, {Jenkins}, {Kjeldsen}, {Koch},
  {Latham}, {Lissauer}, {Marcy}, {Rowe}, {Sasselov}, {Seager}, {Steffen},
  {Torres}, {Basri}, {Brown}, {Charbonneau}, {Christiansen}, {Clarke},
  {Cochran}, {Dupree}, {Fabrycky}, {Fischer}, {Ford}, {Fortney}, {Girouard},
  {Holman}, {Johnson}, {Isaacson}, {Klaus}, {Machalek}, {Moorehead},
  {Morehead}, {Ragozzine}, {Tenenbaum}, {Twicken}, {Quinn}, {VanCleve},
  {Walkowicz}, {Welsh}, {Devore}, \& {Gould}}]{batalha.2011.kfrpk}
{Batalha}, N.~M., {Borucki}, W.~J., {Bryson}, S.~T.,  {et~al.} 2011, \apj, 729,
  27

\bibitem[{{Bean} {et~al.}(2011){Bean}, {D{\'e}sert}, {Kabath}, {Stalder},
  {Seager}, {Miller-Ricci Kempton}, {Berta}, {Homeier}, {Walsh}, \&
  {Seifahrt}}]{bean.2011.ontsspgfema}
{Bean}, J.~L., {D{\'e}sert}, J.-M., {Kabath}, P.,  {et~al.} 2011, accepted to
  \apj, (arXiv:1109.0582)

\bibitem[{{Bean} {et~al.}(2010){Bean}, {Miller-Ricci Kempton}, \&
  {Homeier}}]{bean.2010.gtsse1}
{Bean}, J.~L., {Miller-Ricci Kempton}, E., \& {Homeier}, D. 2010, \nat, 468,
  669

\bibitem[{{Berta} {et~al.}(2011){Berta}, {Charbonneau}, {Bean}, {Irwin},
  {Burke}, {D{\'e}sert}, {Nutzman}, \& {Falco}}]{berta.2011.gsssvtsap}
{Berta}, Z.~K., {Charbonneau}, D., {Bean}, J.,  {et~al.} 2011, \apj, 736, 12

\bibitem[{{Brown}(2001)}]{brown.2001.tsdegpa}
{Brown}, T.~M. 2001, \apj, 553, 1006

\bibitem[{{Brown} {et~al.}(2001){Brown}, {Charbonneau}, {Gilliland}, {Noyes},
  \& {Burrows}}]{brown.2001.hsttptp2}
{Brown}, T.~M., {Charbonneau}, D., {Gilliland}, R.~L., {Noyes}, R.~W., \&
  {Burrows}, A. 2001, \apj, 552, 699

\bibitem[{{Burke} {et~al.}(2010){Burke}, {McCullough}, {Bergeron}, {Long},
  {Gilliland}, {Nelan}, {Johns-Krull}, {Valenti}, \&
  {Janes}}]{burke.2010.notjx}
{Burke}, C.~J., {McCullough}, P.~R., {Bergeron}, L.~E.,  {et~al.} 2010, \apj,
  719, 1796

\bibitem[{{Burke} {et~al.}(2007){Burke}, {McCullough}, {Valenti},
  {Johns-Krull}, {Janes}, {Heasley}, {Summers}, {Stys}, {Bissinger}, {Fleenor},
  {Foote}, {Garc{\'{\i}}a-Melendo}, {Gary}, {Howell}, {Mallia}, {Masi},
  {Taylor}, \& {Vanmunster}}]{burke.2007.xtjmcpmb}
{Burke}, C.~J., {McCullough}, P.~R., {Valenti}, J.~A.,  {et~al.} 2007, \apj,
  671, 2115

\bibitem[{{Carter} {et~al.}(2009){Carter}, {Winn}, {Gilliland}, \&
  {Holman}}]{carter.2009.ntpe1}
{Carter}, J.~A., {Winn}, J.~N., {Gilliland}, R., \& {Holman}, M.~J. 2009, \apj,
  696, 241

\bibitem[{{Carter} {et~al.}(2011){Carter}, {Winn}, {Holman}, {Fabrycky},
  {Berta}, {Burke}, \& {Nutzman}}]{carter.2011.tlcpxsts1}
{Carter}, J.~A., {Winn}, J.~N., {Holman}, M.~J.,  {et~al.} 2011, \apj, 730, 82

\bibitem[{{Carter} {et~al.}(2008){Carter}, {Yee}, {Eastman}, {Gaudi}, \&
  {Winn}}]{carter.2008.aatlouc}
{Carter}, J.~A., {Yee}, J.~C., {Eastman}, J., {Gaudi}, B.~S., \& {Winn}, J.~N.
  2008, \apj, 689, 499

\bibitem[{{Charbonneau} {et~al.}(2009){Charbonneau}, {Berta}, {Irwin}, {Burke},
  {Nutzman}, {Buchhave}, {Lovis}, {Bonfils}, {Latham}, {Udry}, {Murray-Clay},
  {Holman}, {Falco}, {Winn}, {Queloz}, {Pepe}, {Mayor}, {Delfosse}, \&
  {Forveille}}]{charbonneau.2009.stnls}
{Charbonneau}, D., {Berta}, Z.~K., {Irwin}, J.,  {et~al.} 2009, \nat, 462, 891

\bibitem[{{Charbonneau} {et~al.}(2002){Charbonneau}, {Brown}, {Noyes}, \&
  {Gilliland}}]{charbonneau.2002.depa}
{Charbonneau}, D., {Brown}, T.~M., {Noyes}, R.~W., \& {Gilliland}, R.~L. 2002,
  \apj, 568, 377

\bibitem[{{Charbonneau} {et~al.}(2008){Charbonneau}, {Knutson}, {Barman},
  {Allen}, {Mayor}, {Megeath}, {Queloz}, \& {Udry}}]{charbonneau.2008.biese1}
{Charbonneau}, D., {Knutson}, H.~A., {Barman}, T.,  {et~al.} 2008, \apj, 686,
  1341

\bibitem[{{Claret}(2000)}]{claret.2000.nlsamcl2t5ssg}
{Claret}, A. 2000, \aap, 363, 1081

\bibitem[{{Claret} \& {Hauschildt}(2003)}]{claret.2003.lssnmamss}
{Claret}, A., \& {Hauschildt}, P.~H. 2003, \aap, 412, 241

\bibitem[{{Croll} {et~al.}(2011){Croll}, {Albert}, {Jayawardhana},
  {Miller-Ricci Kempton}, {Fortney}, {Murray}, \&
  {Neilson}}]{croll.2011.btss1smmwa}
{Croll}, B., {Albert}, L., {Jayawardhana}, R.,  {et~al.} 2011, \apj, 736, 78

\bibitem[{{Crossfield} {et~al.}(2011){Crossfield}, {Barman}, \&
  {Hansen}}]{crossfield.2011.hdnts1}
{Crossfield}, I.~J.~M., {Barman}, T., \& {Hansen}, B.~M.~S. 2011, \apj, 736,
  132

\bibitem[{{Deming} {et~al.}(2006){Deming}, {Harrington}, {Seager}, \&
  {Richardson}}]{deming.2006.siefep1}
{Deming}, D., {Harrington}, J., {Seager}, S., \& {Richardson}, L.~J. 2006,
  \apj, 644, 560

\bibitem[{{Deming} {et~al.}(2009){Deming}, {Seager}, {Winn}, {Miller-Ricci},
  {Clampin}, {Lindler}, {Greene}, {Charbonneau}, {Laughlin}, {Ricker},
  {Latham}, \& {Ennico}}]{deming.2009.dctseuatsfjwst}
{Deming}, D., {Seager}, S., {Winn}, J.,  {et~al.} 2009, \pasp, 121, 952

\bibitem[{{D{\'e}sert} {et~al.}(2011{\natexlab{a}}){D{\'e}sert}, {Bean},
  {Miller-Ricci Kempton}, {Berta}, {Charbonneau}, {Irwin}, {Fortney}, {Burke},
  \& {Nutzman}}]{desert.2011.oemasg}
{D{\'e}sert}, J.-M., {Bean}, J., {Miller-Ricci Kempton}, E.,  {et~al.}
  2011{\natexlab{a}}, \apjl, 731, L40+

\bibitem[{{D{\'e}sert} {et~al.}(2009){D{\'e}sert}, {Lecavelier des Etangs},
  {H{\'e}brard}, {Sing}, {Ehrenreich}, {Ferlet}, \&
  {Vidal-Madjar}}]{desert.2009.scmate1}
{D{\'e}sert}, J.-M., {Lecavelier des Etangs}, A., {H{\'e}brard}, G.,  {et~al.}
  2009, \apj, 699, 478

\bibitem[{{D{\'e}sert} {et~al.}(2011{\natexlab{b}}){D{\'e}sert}, {Sing},
  {Vidal-Madjar}, {H{\'e}brard}, {Ehrenreich}, {Lecavelier Des Etangs},
  {Parmentier}, {Ferlet}, \& {Henry}}]{desert.2011.tse1isom}
{D{\'e}sert}, J.-M., {Sing}, D., {Vidal-Madjar}, A.,  {et~al.}
  2011{\natexlab{b}}, \aap, 526, A12+

\bibitem[{{Dressel} {et~al.}(2010){Dressel}, {Wong}, {Pavlovsky}, {Long},
  {et~al.}}]{dressel.2010.wfcihv}
{Dressel}, L., {Wong}, M., {Pavlovsky}, C., {Long}, K., {et~al.} 2010, "Wide
  Field Camera 3 Instrument Handbook", Version 2.1, (Baltimore: STScI)

\bibitem[{{Dunkley} {et~al.}(2005){Dunkley}, {Bucher}, {Ferreira}, {Moodley},
  \& {Skordis}}]{dunkley.2005.frmcmctcpe}
{Dunkley}, J., {Bucher}, M., {Ferreira}, P.~G., {Moodley}, K., \& {Skordis}, C.
  2005, \mnras, 356, 925

\bibitem[{{Eastman} {et~al.}(2010){Eastman}, {Siverd}, \&
  {Gaudi}}]{eastman.2010.abtmahbjd}
{Eastman}, J., {Siverd}, R., \& {Gaudi}, B.~S. 2010, \pasp, 122, 935

\bibitem[{{Ehrenreich} {et~al.}(2006){Ehrenreich}, {Tinetti}, {Lecavelier Des
  Etangs}, {Vidal-Madjar}, \& {Selsis}}]{ehrenreich.2006.tsetp}
{Ehrenreich}, D., {Tinetti}, G., {Lecavelier Des Etangs}, A., {Vidal-Madjar},
  A., \& {Selsis}, F. 2006, \aap, 448, 379
\bibitem[{{Encrenaz}(2005)}]{encrenaz.2005.nagpocm}
{Encrenaz}, T. 2005, \ssr, 116, 99

\bibitem[{{Ford}(2005)}]{ford.2005.quoep}
{Ford}, E.~B. 2005, \aj, 129, 1706

\bibitem[{{Fortney}(2005)}]{fortney.2005.ecctpawts}
{Fortney}, J.~J. 2005, \mnras, 364, 649

\bibitem[{{Gautier} {et~al.}(1995){Gautier}, {Conrath}, {Owen}, {de Pater}, \&
  {Atreya}}]{gautier.1995.tn}
{Gautier}, D., {Conrath}, B.~J., {Owen}, T., {de Pater}, I., \& {Atreya}, S.~K.
  1995, in Neptune and Triton, ed. {D.~P.~Cruikshank, M.~S.~Matthews, \&
  A.~M.~Schumann}, 547--611

\bibitem[{{Gibson} {et~al.}(2011{\natexlab{a}}){Gibson}, {Pont}, \&
  {Aigrain}}]{gibson.2011.lnts1gxcemf}
{Gibson}, N.~P., {Pont}, F., \& {Aigrain}, S. 2011{\natexlab{b}}, \mnras, 411,
  2199

\bibitem[{{Gibson} {et~al.}(2011{\natexlab{b}}){Gibson}, {Aigrain}, {Roberts},
  {Evans}, {Osborne}, \& {Pont}}]{gibson.2011.gpfmisats}
{Gibson}, N.~P., {Aigrain}, S., {Roberts}, S.,  {et~al.} 2011{\natexlab{a}}, accepted to MNRAS, (arXiv:1109.3251)

\bibitem[{{Gregory}(2005)}]{gregory.2005.bldapscawms}
{Gregory}, P.~C. 2005, {Bayesian Logical Data Analysis for the Physical
  Sciences: A Comparative Approach with `Mathematica' Support}, ed. {Gregory,
  P.~C.} (Cambridge University Press)

\bibitem[{{Guillot} \& {Gautier}(2009)}]{guillot.2009.gp}
{Guillot}, T., \& {Gautier}, D. 2009, arXiv:0912.2019

\bibitem[{{Hauschildt} {et~al.}(1999){Hauschildt}, {Allard}, \&
  {Baron}}]{hauschildt.1999.nmag3}
{Hauschildt}, P.~H., {Allard}, F., \& {Baron}, E. 1999, \apj, 512, 377

\bibitem[{{Helling} {et~al.}(2008){Helling}, {Woitke}, \&
  {Thi}}]{helling.2008.dbdepiccsaqcl}
{Helling}, C., {Woitke}, P., \& {Thi}, W.-F. 2008, \aap, 485, 547

\bibitem[{{Hilbert} \& {McCullough}(2011)}]{hilbert.2011.iccmmo}
{Hilbert}, B., \& {McCullough}, P. 2011, {WFC3 Instrument Science Report 2011-10 (Baltimore, MD: STScI)}

\bibitem[{{Hogg} {et~al.}(2010){Hogg}, {Bovy}, \& {Lang}}]{hogg.2010.darfmd}
{Hogg}, D.~W., {Bovy}, J., \& {Lang}, D. 2010, arXiv:1008.4686

\bibitem[{{Holman} {et~al.}(2006){Holman}, {Winn}, {Latham}, {O'Donovan},
  {Charbonneau}, {Bakos}, {Esquerdo}, {Hergenrother}, {Everett}, \&
  {P{\'a}l}}]{holman.2006.tlcpifctex}
{Holman}, M.~J., {Winn}, J.~N., {Latham}, D.~W.,  {et~al.} 2006, \apj, 652,
  1715

\bibitem[{{Hubbard} {et~al.}(2001){Hubbard}, {Fortney}, {Lunine}, {Burrows},
  {Sudarsky}, \& {Pinto}}]{hubbard.2001.tegpt}
{Hubbard}, W.~B., {Fortney}, J.~J., {Lunine}, J.~I.,  {et~al.} 2001, \apj, 560,
  413
  
\bibitem[{{Irwin} {et~al.}(2011){Irwin}, {Quinn}, {Berta}, {Latham}, {Torres},
  {Burke}, {Charbonneau}, {Dittmann}, {Esquerdo}, {Stefanik}, {Oksanen},
  {Buchhave}, {Nutzman}, {Berlind}, {Calkins}, \&
  {Falco}}]{irwin.2011.ljd4mebfmts}
{Irwin}, J.~M., {Quinn}, S.~N., {Berta}, Z.~K.,  {et~al.} 2011, accepted to \apj, (arXiv:1109.2055)

\bibitem[{{Kaltenegger} \& {Traub}(2009)}]{kaltenegger.2009.tep}
{Kaltenegger}, L., \& {Traub}, W.~A. 2009, \apj, 698, 519

\bibitem[{{Kim Quijano} {et~al.}(2009)}]{kim-quijano.2009.wmh}
{Kim Quijano}, J., {et~al.} 2009, ``WFC3 Mini-Data Handbook'', Version 1.0,
  (Baltimore: STScI)

\bibitem[{{Kipping}(2009)}]{kipping.2009.ttee}
{Kipping}, D.~M. 2009, \mnras, 392, 181

\bibitem[{{Kipping}(2011)}]{kipping.2011.lagdpt}
{Kipping}, D.~M. 2011, \mnras, 416, 689

\bibitem[{{Knutson} {et~al.}(2007{\natexlab{a}}){Knutson}, {Charbonneau},
  {Allen}, {Fortney}, {Agol}, {Cowan}, {Showman}, {Cooper}, \&
  {Megeath}}]{knutson.2007.dcep1}
{Knutson}, H.~A., {Charbonneau}, D., {Allen}, L.~E.,  {et~al.}
  2007{\natexlab{a}}, \nat, 447, 183

\bibitem[{{Knutson} {et~al.}(2007{\natexlab{b}}){Knutson}, {Charbonneau},
  {Noyes}, {Brown}, {Gilliland}, {Knutson}, {Charbonneau}, {Noyes}, {Brown},
  {Gilliland}, {Knutson}, {Charbonneau}, {Noyes}, {Brown}, \&
  {Gilliland}}]{knutson.2007.uslrp2}
{Knutson}, H.~A., {Charbonneau}, D., {Noyes}, R.~W.,  {et~al.}
  2007{\natexlab{b}}, \apj, 655, 564

\bibitem[{{Knutson} {et~al.}(2011){Knutson}, {Madhusudhan}, {Cowan},
  {Christiansen}, {Agol}, {Deming}, {Desert}, {Charbonneau}, {Henry},
  {Homeier}, {Langton}, {Laughlin}, \& {Seager}}]{knutson.2011.stse4esvcdfv}
{Knutson}, H.~A., {Madhusudhan}, N., {Cowan}, N.~B.,  {et~al.} 2011, accepted
  to \apj, (arXiv:1104.2901)

\bibitem[{{K{\"u}mmel} {et~al.}(2011){K{\"u}mmel}, {Kuntschner}, {Walsh}, \&
  {Bushouse}}]{kummel.2011.miwggg}
{K{\"u}mmel}, M., {Kuntschner}, H., {Walsh}, J.~R., \& {Bushouse}, H. 2011,
  {Instrument Science Report WFC3-2011-01 (Baltimore, MD: STScI)}

\bibitem[{{K{\"u}mmel} {et~al.}(2010){K{\"u}mmel}, {Walsh}, \&
  {Kuntschner}}]{kummel.2010.umv}
{K{\"u}mmel}, M., {Walsh}, J.~R., \& {Kuntschner}, H. 2010, "aXe User Manual",
  Version 2.1, (Baltimore: STScI)

\bibitem[{{K{\"u}mmel} {et~al.}(2009){K{\"u}mmel}, {Walsh}, {Pirzkal},
  {Kuntschner}, \& {Pasquali}}]{kummel.2009.ssdes}
{K{\"u}mmel}, M., {Walsh}, J.~R., {Pirzkal}, N., {Kuntschner}, H., \&
  {Pasquali}, A. 2009, \pasp, 121, 59

\bibitem[{{Kundurthy} {et~al.}(2011){Kundurthy}, {Agol}, {Becker}, {Barnes},
  {Williams}, \& {Mukadam}}]{kundurthy.2011.ao1spesa}
{Kundurthy}, P., {Agol}, E., {Becker}, A.~C.,  {et~al.} 2011, \apj, 731, 123

\bibitem[{{Kuntschner} {et~al.}(2009){Kuntschner}, {Bushouse}, {K{\"u}mmel}, \&
  {Walsh}}]{kuntschner.2009.wsp1cgg}
{Kuntschner}, H., {Bushouse}, H., {K{\"u}mmel}, M., \& {Walsh}, J.~R. 2009,
  {ST-ECF Instrument Science Report WFC3-2009-17 (Baltimore, MD: STScI)}

\bibitem[{{Kuntschner} {et~al.}(2008){Kuntschner}, {Bushouse}, {Walsh}, \&
  {K{\"u}mmel}}]{kuntschner.2008.gcwg}
{Kuntschner}, H., {Bushouse}, H., {Walsh}, J.~R., \& {K{\"u}mmel}, M. 2008,
  {ST-ECF Instrument Science Report WFC3-2008-16 (Baltimore, MD: STScI)}

\bibitem[{{Kuntschner} {et~al.}(2011){Kuntschner}, {K{\"u}mmel}, {Walsh}, \&
  {Bushouse}}]{kuntschner.2011.rfcwggg}
{Kuntschner}, H., {K{\"u}mmel}, M., {Walsh}, J.~R., \& {Bushouse}, H. 2011,
  {ST-ECF Instrument Science Report WFC3-2011-05 (Baltimore, MD: STScI)}

\bibitem[{{Lecavelier Des Etangs} {et~al.}(2008){Lecavelier Des Etangs},
  {Pont}, {Vidal-Madjar}, \& {Sing}}]{lecavelier-des-etangs.2008.rsts1}
{Lecavelier Des Etangs}, A., {Pont}, F., {Vidal-Madjar}, A., \& {Sing}, D.
  2008, \aap, 481, L83

\bibitem[{{L{\'e}ger} {et~al.}(2009){L{\'e}ger}, {Rouan}, {Schneider}, {Barge},
  {Fridlund}, {Samuel}, {Ollivier}, {Guenther}, {Deleuil}, {Deeg}, {Auvergne},
  {Alonso}, {Aigrain}, {Alapini}, {Almenara}, {Baglin}, {Barbieri}, {Bruntt},
  {Bord{\'e}}, {Bouchy}, {Cabrera}, {Catala}, {Carone}, {Carpano}, {Csizmadia},
  {Dvorak}, {Erikson}, {Ferraz-Mello}, {Foing}, {Fressin}, {Gandolfi},
  {Gillon}, {Gondoin}, {Grasset}, {Guillot}, {Hatzes}, {H{\'e}brard}, {Jorda},
  {Lammer}, {Llebaria}, {Loeillet}, {Mayor}, {Mazeh}, {Moutou}, {P{\"a}tzold},
  {Pont}, {Queloz}, {Rauer}, {Renner}, {Samadi}, {Shporer}, {Sotin}, {Tingley},
  {Wuchterl}, {Adda}, {Agogu}, {Appourchaux}, {Ballans}, {Baron}, {Beaufort},
  {Bellenger}, {Berlin}, {Bernardi}, {Blouin}, {Baudin}, {Bodin}, {Boisnard},
  {Boit}, {Bonneau}, {Borzeix}, {Briet}, {Buey}, {Butler}, {Cailleau},
  {Cautain}, {Chabaud}, {Chaintreuil}, {Chiavassa}, {Costes}, {Cuna Parrho},
  {de Oliveira Fialho}, {Decaudin}, {Defise}, {Djalal}, {Epstein}, {Exil},
  {Faur{\'e}}, {Fenouillet}, {Gaboriaud}, {Gallic}, {Gamet}, {Gavalda},
  {Grolleau}, {Gruneisen}, {Gueguen}, {Guis}, {Guivarc'h}, {Guterman},
  {Hallouard}, {Hasiba}, {Heuripeau}, {Huntzinger}, {Hustaix}, {Imad},
  {Imbert}, {Johlander}, {Jouret}, {Journoud}, {Karioty}, {Kerjean},
  {Lafaille}, {Lafond}, {Lam-Trong}, {Landiech}, {Lapeyrere}, {Larqu{\'e}},
  {Laudet}, {Lautier}, {Lecann}, {Lefevre}, {Leruyet}, {Levacher}, {Magnan},
  {Mazy}, {Mertens}, {Mesnager}, {Meunier}, {Michel}, {Monjoin}, {Naudet},
  {Nguyen-Kim}, {Orcesi}, {Ottacher}, {Perez}, {Peter}, {Plasson}, {Plesseria},
  {Pontet}, {Pradines}, {Quentin}, {Reynaud}, {Rolland}, {Rollenhagen},
  {Romagnan}, {Russ}, {Schmidt}, {Schwartz}, {Sebbag}, {Sedes}, {Smit},
  {Steller}, {Sunter}, {Surace}, {Tello}, {Tiph{\`e}ne}, {Toulouse}, {Ulmer},
  {Vandermarcq}, {Vergnault}, {Vuillemin}, \&
  {Zanatta}}]{leger.2009.tefcsmvcfswmr}
{L{\'e}ger}, A., {Rouan}, D., {Schneider}, J.,  {et~al.} 2009, \aap, 506, 287

\bibitem[{{Lissauer} {et~al.}(2011){Lissauer}, {Fabrycky}, {Ford}, {Borucki},
  {Fressin}, {Marcy}, {Orosz}, {Rowe}, {Torres}, {Welsh}, {Batalha}, {Bryson},
  {Buchhave}, {Caldwell}, {Carter}, {Charbonneau}, {Christiansen}, {Cochran},
  {Desert}, {Dunham}, {Fanelli}, {Fortney}, {Gautier}, {Geary}, {Gilliland},
  {Haas}, {Hall}, {Holman}, {Koch}, {Latham}, {Lopez}, {McCauliff}, {Miller},
  {Morehead}, {Quintana}, {Ragozzine}, {Sasselov}, {Short}, \&
  {Steffen}}]{lissauer.2011.cpsllptk}
{Lissauer}, J.~J., {Fabrycky}, D.~C., {Ford}, E.~B.,  {et~al.} 2011, \nat, 470,
  53

\bibitem[{{Lodders} \& {Fegley}(2006)}]{lodders.2006.cmso}
{Lodders}, K., \& {Fegley}, Jr., B. 2006, {Chemistry of Low Mass Substellar
  Objects}, ed. {Mason, J.~W.} (Springer Verlag), 1--+

\bibitem[{{Long} {et~al.}(2010){Long}, {Baggett}, {Deustua}, \&
  {Riess}}]{long.2010.wpmcutle}
{Long}, K.~S., {Baggett}, S., {Deustua}, S., \& {Riess}, A. 2010, {WFC3 Instrument Science Report 2010-17 (Baltimore, MD: STScI)}

\bibitem[{{Long} {et~al.}(2011){Long}, {Wheeler}, \&
  {Bushouse}}]{long.2011.dtp}
{Long}, K.~S., {Wheeler}, T., \& {Bushouse}, H. 2011, {WFC3 Instrument Science Report 2011-09 (Baltimore, MD: STScI)}

\bibitem[{{Mandel} \& {Agol}(2002)}]{mandel.2002.alcpts}
{Mandel}, K., \& {Agol}, E. 2002, \apjl, 580, L171

\bibitem[{{Markwardt}(2009)}]{markwardt.2009.nlfwm}
{Markwardt}, C.~B. 2009, in Astronomical Society of the Pacific Conference
  Series, Vol. 411, Astronomical Data Analysis Software and Systems XVIII, ed.
  {D.~A.~Bohlender, D.~Durand, \& P.~Dowler}, 251--+

\bibitem[{{McCullough}(2008)}]{mccullough.2008.icpd}
{McCullough}, P. 2008, {Instrument Science Report WFC3 2008-26 (Baltimore, MD: STScI)}

\bibitem[{{McCullough} \& {Deustua}(2008)}]{mccullough.2008.wtp}
{McCullough}, P., \& {Deustua}, S. 2008, {Instrument Science Report WFC3 2008-33 (Baltimore, MD: STScI)}

\bibitem[{{McCullough} \& {MacKenty}(2011)}]{mccullough.2011.si}
{McCullough}, P.~R., \& {MacKenty}, J.~W. 2011, {WFC Space Telescope Analysis
  Newletter 6 (Baltimore, MD: STScI)}

\bibitem[{{Menou}(2011)}]{menou.2011.accg}
{Menou}, K. 2011, submitted to ApJL, (arXiv:1109.1574)

\bibitem[{{Miller-Ricci} \& {Fortney}(2010)}]{miller-ricci.2010.nats1}
{Miller-Ricci}, E., \& {Fortney}, J.~J. 2010, \apjl, 716, L74

\bibitem[{{Miller-Ricci} {et~al.}(2009){Miller-Ricci}, {Seager}, \&
  {Sasselov}}]{miller-ricci.2009.assdbhha}
{Miller-Ricci}, E., {Seager}, S., \& {Sasselov}, D. 2009, \apj, 690, 1056

\bibitem[{{Miller-Ricci Kempton} {et~al.}(2011){Miller-Ricci Kempton},
  {Zahnle}, \& {Fortney}}]{miller-ricci-kempton.2011.ac1pc}
{Miller-Ricci Kempton}, E., {Zahnle}, K., \& {Fortney}, J.~J. 2011, submitted
  to \apj, (arXiv:1104.5477)

\bibitem[{{Nettelmann} {et~al.}(2011){Nettelmann}, {Fortney}, {Kramm}, \&
  {Redmer}}]{nettelmann.2011.tesmts1}
{Nettelmann}, N., {Fortney}, J.~J., {Kramm}, U., \& {Redmer}, R. 2011, \apj,
  733, 2

\bibitem[{{Nutzman} {et~al.}(2011){Nutzman}, {Gilliland}, {McCullough},
  {Charbonneau}, {Christensen-Dalsgaard}, {Kjeldsen}, {Nelan}, {Brown}, \&
  {Holman}}]{nutzman.2011.peppes1ehstfgstao}
{Nutzman}, P., {Gilliland}, R.~L., {McCullough}, P.~R.,  {et~al.} 2011, \apj,
  726, 3

\bibitem[{{Pall{\'e}} {et~al.}(2011){Pall{\'e}}, {Zapatero Osorio}, \&
  {Garc{\'{\i}}a Mu{\~n}oz}}]{palle.2011.catrpald}
{Pall{\'e}}, E., {Zapatero Osorio}, M.~R., \& {Garc{\'{\i}}a Mu{\~n}oz}, A.
  2011, \apj, 728, 19

\bibitem[{{Pirzkal} {et~al.}(2011){Pirzkal}, {Mack}, {Dahlen}, \&
  {Sabbi}}]{pirzkal.2011.fgiwf}
{Pirzkal}, N., {Mack}, J., {Dahlen}, T., \& {Sabbi}, E. 2011, {Instrument Science Report WFC3-2011-11 (Baltimore, MD: STScI)}

\bibitem[{{Pont} {et~al.}(2007){Pont}, {Gilliland}, {Moutou}, {Charbonneau},
  {Bouchy}, {Brown}, {Mayor}, {Queloz}, {Santos}, \&
  {Udry}}]{pont.2007.hsttppt1mrs}
{Pont}, F., {Gilliland}, R.~L., {Moutou}, C.,  {et~al.} 2007, \aap, 476, 1347

\bibitem[{{Pont} {et~al.}(2008){Pont}, {Knutson}, {Gilliland}, {Moutou}, \&
  {Charbonneau}}]{pont.2008.dahep0mts1wh}
{Pont}, F., {Knutson}, H., {Gilliland}, R.~L., {Moutou}, C., \& {Charbonneau},
  D. 2008, \mnras, 385, 109

\bibitem[{{Rogers} {et~al.}(2011){Rogers}, {Bodenheimer}, {Lissauer}, \&
  {Seager}}]{rogers.2011.fsle}
{Rogers}, L.~A., {Bodenheimer}, P., {Lissauer}, J.~J., \& {Seager}, S. 2011,
  \apj, 738, 59

\bibitem[{{Rogers} \& {Seager}(2010)}]{rogers.2010.tpol1}
{Rogers}, L.~A., \& {Seager}, S. 2010, \apj, 716, 1208

\bibitem[{{Sada} {et~al.}(2010){Sada}, {Deming}, {Jackson}, {Jennings},
  {Peterson}, {Haase}, {Bays}, {O'Gorman}, \& {Lundsford}}]{sada.2010.rtse1}
{Sada}, P.~V., {Deming}, D., {Jackson}, B.,  {et~al.} 2010, \apjl, 720, L215

\bibitem[{{Seager} \& {Sasselov}(2000)}]{seager.2000.ttsdegpt}
{Seager}, S., \& {Sasselov}, D.~D. 2000, \apj, 537, 916
\bibitem[{{Seager} {et~al.}(2007){Seager}, {Kuchner}, {Hier-Majumder}, \&
  {Militzer}}]{seager.2007.mrse}
{Seager}, S., {Kuchner}, M., {Hier-Majumder}, C.~A., \& {Militzer}, B. 2007,
  \apj, 669, 1279

\bibitem[{{Sing} {et~al.}(2009){Sing}, {D{\'e}sert}, {Lecavelier Des Etangs},
  {Ballester}, {Vidal-Madjar}, {Parmentier}, {Hebrard}, \&
  {Henry}}]{sing.2009.tse1iswfhwn}
{Sing}, D.~K., {D{\'e}sert}, J.-M., {Lecavelier Des Etangs}, A.,  {et~al.}
  2009, \aap, 505, 891

\bibitem[{{Sing} {et~al.}(2011){Sing}, {Pont}, {Aigrain}, {Charbonneau},
  {D{\'e}sert}, {Gibson}, {Gilliland}, {Hayek}, {Henry}, {Knutson}, {Lecavelier
  Des Etangs}, {Mazeh}, \& {Shporer}}]{sing.2011.hsttse1hahonws}
{Sing}, D.~K., {Pont}, F., {Aigrain}, S.,  {et~al.} 2011, \mnras, 1159

\bibitem[{Sivia(1996)}]{sivia.1996.dabt}
Sivia, D.~S. 1996, Data Analysis: A Bayesian Tutorial (Oxford University
  Press), xi, 189 p.

\bibitem[{{Smith} {et~al.}(2008{\natexlab{a}}){Smith}, {Zavodny}, {Rahmer}, \&
  {Bonati}}]{smith.2008.tiphp}
{Smith}, R.~M., {Zavodny}, M., {Rahmer}, G., \& {Bonati}, M.
  2008{\natexlab{a}}, in Society of Photo-Optical Instrumentation Engineers
  (SPIE) Conference Series, Vol. 7021, Society of Photo-Optical Instrumentation
  Engineers (SPIE) Conference Series

\bibitem[{{Smith} {et~al.}(2008{\natexlab{b}}){Smith}, {Zavodny}, {Rahmer}, \&
  {Bonati}}]{smith.2008.ciphp}
{Smith}, R.~M., {Zavodny}, M., {Rahmer}, G., \& {Bonati}, M.
  2008{\natexlab{b}}, in Society of Photo-Optical Instrumentation Engineers
  (SPIE) Conference Series, Vol. 7021, Society of Photo-Optical Instrumentation
  Engineers (SPIE) Conference Series

\bibitem[{{Southworth}(2008)}]{southworth.2008.hstepila}
{Southworth}, J. 2008, \mnras, 386, 1644

\bibitem[{{Swain} {et~al.}(2008){Swain}, {Vasisht}, \&
  {Tinetti}}]{swain.2008.pmaep}
{Swain}, M.~R., {Vasisht}, G., \& {Tinetti}, G. 2008, \nat, 452, 329

\bibitem[{{van Altena} {et~al.}(1995){van Altena}, {Lee}, \&
  {Hoffleit}}]{van-altena.1995.gctsp}
{van Altena}, W.~F., {Lee}, J.~T., \& {Hoffleit}, E.~D. 1995, {The general
  catalogue of trigonometric [stellar] parallaxes}, ed. {van Altena, W.~F.,
  Lee, J.~T., \& Hoffleit, E.~D.}

\bibitem[{{van Hamme}(1993)}]{van-hamme.1993.lcmbslc}
{van Hamme}, W. 1993, \aj, 106, 2096

\bibitem[{{Viana} \& {Baggett}(2010)}]{viana.2010.wtc}
{Viana}, A.~C., \& {Baggett}, S. 2010, {WFC3 TV3 Testing: IR Crosstalk}, Tech.
  rep.

\bibitem[{{Winn} {et~al.}(2011){Winn}, {Matthews}, {Dawson}, {Fabrycky},
  {Holman}, {Kallinger}, {Kuschnig}, {Sasselov}, {Dragomir}, {Guenther},
  {Moffat}, {Rowe}, {Rucinski}, \& {Weiss}}]{winn.2011.stns}
{Winn}, J.~N., {Matthews}, J.~M., {Dawson}, R.~I.,  {et~al.} 2011, \apjl, 737,
  L18+

\end{thebibliography}
\end{document}